\documentstyle[epsfig,graphics,a4wide,aps]{revtex} 

\draft

\begin{document}

\title{The XY-model with $Z_2$ Symmetry:\\ Finite-size scaling analysis
  using the Broad Histogram Method} 
\author{J.D. Mu\~noz$^{1,2,}$\footnote{Corresponding author:
    jdmunoz@ica1.uni-stuttgart.de} and A.R. Lima$^{3,4}$}
\address{(1) Institute for Computer Applications 1, Stuttgart
  University\\Pfaffenwaldring 27, 70569, Stuttgart, Germany\\(2) Permanent
  address: Dpto de F\'{\i}sica, Universidad Nacional de Colombia\\Bogot\'a
  D.C., Colombia\\(3) PMMH, \'Ecole Sup\'erieure de Physique et Chimie
  Industrielles (ESPCI)\\ 10, rue Vauquelin, 75231 Paris, 
  Cedex 05, France \\ (4) Instituto de F\'{\i}sica, Universidade
  Federal Fluminense \\ Av. Litor\^anea, s/n, 24210-340 Niter\'oi, RJ,
  Brazil}
\date{\today}
\maketitle

\begin{abstract}
  In this work we investigate the classical ferromagnetic XY-model in
  two dimensions subject to a symmetry breaking field which impose a
  $Z_2$ symmetry to the system.  We used the broad histogram method
  combined with microcanonical simulations and finite-size scaling
  analysis to estimate the critical temperature and critical exponents
  of this system with little computational effort.  In addition, we
  present a general procedure which makes possible to use the broad
  histogram method for continuous systems, also when the macroscopic
  quantities needed by the method cannot be obtained analytically.
  Our results are robust under the choice of four different
  pseudo-random number generators.
\end{abstract}

\section{ Introduction }
\label{sec:intro}
%
Consider a XY-model with Hamiltonian
\begin{equation}
  \label{Hamiltonian}
{\cal H} = -J\sum_{\langle ij \rangle}{\cos(\theta_i - \theta_j)} -
h_p\sum_{i}{\cos(p\theta_i)} \; ,
\end{equation} 
where $\theta_i$ denotes the angle between the spin $\vec{\sigma}_i$, in
site $i$, and the $x$-axis, $J$ is a coupling constant, the first
summation extends over all pairs of nearest neighbors and the second
one over all sites.  The second term in (\ref{Hamiltonian})
corresponds to the energy associated with a symmetry breaking field,
characterized by a positive integer $p$ and a constant
$h_p$.  This model is important since it may describe many real
magnetic crystals subject to symmetry-breaking crystalline fields
\cite{jongh74}. Experimental crystalline fields corresponding to
$p=2$, $4$ and $6$ are expected.  For $p=1$ the model corresponds to a
XY-model with an external magnetic field. For $p=2$ we have a system
with $Z_2$ symmetry (which we call XY2).  This latter case
is specially interesting because: (i) It was studied by Jos\'e {\em
  et.  al.} who were unable to ascertain the small $h_2$ structure of
the phase diagram\cite{jose77}, (ii) The critical
temperature $T_{\rm c}$ for this system, neither for $h_2=1$, 
is not known exactly.
(iii) Recently, Resende and Costa \cite{resende98} have founded
differences of 2\% in the value of $T_{\rm c}$ 
depending on the pseudo-random number generator
employed, i.e., $T_{\rm c}=1.529(5)$ with {\textsc RAN2} \cite{press92}
or $T_{\rm c}=1.491(6)$ with {\textsc RANLUX} in the fourth {\it
  luxury} level \cite{james94},
and (iv) the quantities needed to implement the recently introduced
``Broad Histogram Method'' (BHM) \cite{pmco96,pmco98a} 
for this system are hard to compute.

In this paper we have two main aims: (i) to show how to implement the BHM
when there are no analytical expressions for its quantities and, (ii) to
use the BHM, together with microcanonical simulations, 
to calculate the critical temperature and the critical
exponents for the XY2 model ($h_2=1$).

In the next section we review the implementation of the BHM for
continuous systems and we describe how to apply it to the XY2 model 
and, in extend, to {\bf any} continuous model.  
In section (\ref{sec:implement}) some details 
on the sampling process employed are presented. 
Section (\ref{sec:results}) shows
the results of the finite-size scaling analysis.
Finally, we summarize some comments and conclusion
in section (\ref{sec:conclusion}).

\section{The Broad Histogram Method}
\label{sec:bhm}
%
The broad histogram method (BHM) \cite{pmco96,pmco98a} is a novel analysis
technique introduced in 1996 by de Oliveira {\em et. al.} in
order to calculate the spectral degeneracy $g(E)$ of a given system. 
Instead of investigating the form of the histogram of visits, the new
method computes $g(E)$ from the microcanonical averages of a special
macroscopic quantity $N({\bf X}, \Delta E)$ defined below.
This new analysis method can be used in conjunction with any sampling strategy
used previously (microcanonical,canonical, multicanonical, etc.). 
Since the basic quantities are  microcanonical averages, 
the method does not depend on the form of histogram of visits
and sample sets of different runs can be joined together 
just by cummulating them in the same histograms.
Furthermore, each sample contributes with a number of several digits
to these averages, a much richer information that just increasing
by unity some entry in the histogram of visits.
As a result, the BHM can estimate thermodynamic averages with more precision
on broader temperature ranges than other reweighting methods \cite{lima00,kastner00}.

Consider some symmetric relationship $R$ connecting pairs of states 
in the phase space of the system
and define a new macroscopic quantity $N({\bf X}, \Delta E)$
as the number of connections starting from state ${\bf X}$ that changes the energy
of this state by $\Delta E$.
Now, consider the subspaces $\Gamma (E)$ and $\Gamma (E+\Delta E)$ of 
all states with energy $E$ and $E+\Delta E$, respectively.
Since $R$ is symmetric, the number of connections going from $\Gamma (E)$ to
$\Gamma (E+\Delta E)$ equals the number of connections coming back from 
$\Gamma (E+\Delta E)$ to $\Gamma (E)$, or
\begin{equation}
  \label{bhrel}
  g(E)\langle N(E, \Delta E)\rangle   =  g(E+\Delta E)\langle
  N(E+\Delta E, -\Delta E)\rangle \; ,
\end{equation}
where $g(E)$ is the number of microstates with energy $E$ and 
$\langle N(E, \Delta E)\rangle$ is the average of $N({\bf X}, \Delta E)$
on $\Gamma (E)$.
This is the so called {\it broad histogram equation}.
It means that the ratio $g(E+\Delta E)/g(E)$ can be obtained by
estimation of the averages $\langle N(E, \Delta E)\rangle$.
Equation (\ref{bhrel}) is completely general and is exact for any
statistical model \cite{pmco98d}.
Starting from an initial value $g(E_0)$ and applying the method iteratively,
the density of states function $g(E)$ can be obtained 
in steps of $\Delta E$ on a whole energy range.

By taking logarithm at both sides of eq. (\ref{bhrel}) we arrive to the 
following expression for $\beta(E)$$:=$$d \ln{[g(E)]} / dE$:
\begin{eqnarray}
  \label{beta_BHM}
  \beta (E) \simeq {1 \over \Delta E} \ln \left( 
        { \langle N(E, \Delta E)\rangle
        \over
          \langle N(E+\Delta E, -\Delta E)\rangle
        } \right) \; .
\end{eqnarray}
Then, an alternative procedure is to use this expression to estimate $\beta (E)$ 
at several energy values and integrate it
to obtain an approximation of $S(E)$$:=$$\ln{[g(E)]}$, the microcanonical entropy.

The BHM is an analysis procedure on a set of samples obtained previously.
The samples itself can be colleted with every sampling method avaible as,
for instance, microcanonical\cite{care96,creutz83}, canonical\cite{metropolis53,glauber63} 
(including cluster \cite{wang90,wolff89} algorithms)
or multicanonical\cite{berg91,berg92,berg93,berg96,lee93,lee94} and $1/k$ \cite{hesselbo95} procedures, 
and some of them have been already tested 
\cite{pmco96,pmco98a,lima00,kastner00,pmco98b,wang99b,wang99c,munoz98,munoz99,munoz99b,lima99b,lima00b}.
The averages $\langle N(E, \Delta E)\rangle$ can be estimated just by
cumulating the quantities $N({\bf X}, \Delta E)$ on histograms in $E$ 
and by dividing them into the number of samples at each energy bin, 
so far states with the same energy are taken into the sample with the same probability
(This is the usual procedure to compute microcanonical averages).

If we are interested in canonical averages $\langle Q \rangle_T$ at temperature $T$
of a certain quantity $Q$, we need an additional histogram in $E$ to
cumulate $Q(X)$ for all states in the sample and compute the 
microcanonical averages $\langle Q(E) \rangle$.
Once $g(E)$ and $\langle Q(E) \rangle$ have been determined, 
the canonical aveage $\langle Q \rangle_T$ of a certain quantity $Q$
are computed as
\begin{eqnarray}
  \langle Q \rangle_T&=&{1 \over Z(T)} \sum_E \langle Q(E) \rangle \, 
                                        g(E) \exp[-E/k_{\rm B}T]  
  \nonumber \\
  Z(T)&=& \sum_E g(E) \exp[-E/k_{\rm B}T] \; 
  \label{meanQE}
\end{eqnarray}
for any desired temperature, where $k_B$ is the Boltzmann constant.

The BHM method was originally used to investigate discrete models but,  
recently, Mu\~noz and Herrmannn have shown how to extend it
for continuous systems, as the XY \cite{munoz98,munoz99b} and Heisenberg 
\cite{munoz99} models. The relationship $R$ is replaced by a symmetric
transition probability $R({\bf Y}|{\bf X})=R({\bf X}|{\bf Y})$, and the 
number of connections $N({\bf X}, \Delta E)$ is redefined 
as the probability, $f_{\Delta E}({\bf X}, \Delta E)$, to obtain an
energy change between $\Delta E$ and $\Delta E + d\Delta E$ with this
transition probability $R$.
Hence, the problem reduces to that of calculating the 
probability density function (p.d.f.)
$f_{\Delta E}({\bf X}, \Delta E)$ for a given configuration ${\bf X}$.

This calculation can be easily done for the XY and Heisenberg
models, where a partial analytical computation is easy to perform. 
For the XY model with Z2 symmetry the problem is
more difficult. Consider the transition probability $R$ defined by
\begin{eqnarray}
  R({\bf Y}|{\bf X})=\left\{ \begin{array}{cl}
        {1 \over 2 \pi N} & \mbox{ if ${\bf Y}$ can be reach from ${\bf X}$ by a} \\
                          & \mbox{ single-spin turn} \\
        0                 & \mbox{ otherwise}
         \end{array} \right. \; .
  \label{R_XY}
\end{eqnarray}
Here, $N$ is the number of sites.  
This transition probability corresponds to the following 
protocol of virtual random movements: 
first, a site $i$ is chosen at random (with probability $1/N$); 
then, a new angle $\theta_{i \; {\rm new}}$ is chosen at random on the 
interval $[-\pi,\pi)$. Thus, $\theta_{i \; {\rm new}}$ is a random variable 
with p.d.f.
\begin{equation}
f_{\theta_{i \; {\rm new}}}(\theta) = \left\{ \begin{array}{cl} 
  1/2\pi & \mbox{ : if } \pi < \theta_{i \; {\rm new}} < \pi \\
  0      &\mbox{ : otherwise}
  \end{array} \right. \; .
\end{equation}

If the spin $i$ was chosen to be turn, its bonds and interaction with the external
field are the only ones that change the energy value. Let us define the energy
associated to spin $i$ as
\begin{eqnarray}
  \frac{\epsilon_i}{J} &=& -\sum_j{\cos(\theta_i-\theta_j)} - \frac{h}{J}\cos(2\theta_i) \nonumber \\
  &=& -A_{i}\cos(\theta_i) - B_{i}\sin(\theta_i) - \frac{h}{J}\cos(2\theta_i) \nonumber \\
  &\equiv& g(A_{i}, B_{i}, \theta_{i}) \; ,
  \label{epsiloni}
\end{eqnarray}
where $A_i$$=$$\sum_{j}{\cos(\theta_j)}$, $B_i$$=$$
\sum_{j}{\sin(\theta_j)}$ and the summations run over all nearest neighbors of
spin $i$.  If $\theta_i$ changes to $\theta_{i \; {\rm new}}$
the energy $\epsilon_i$ changes to  $\epsilon_{i \; {\rm new}}$.
What is the p.d.f. of this random variable?
To find the p.d.f. $f_{\epsilon_{i}^{\rm new}}(\epsilon)$ for a specific
$\epsilon$ we need to solve the equation
\begin{equation}
g(A_{i}, B_{i}, \theta)-\epsilon=0.
\label{eq:roots}
\end{equation}
Denoting its real roots by $\theta_1, \theta_2, ...,\theta_n$, the
p.d.f. is written as \cite{papoulis83}
\begin{equation}
  f_{\epsilon_{i}^{\rm new}}(\epsilon) = \frac{f_{\theta_{i \; {\rm
        new}}}(\theta_1)}{|dg/d\theta(A_{i}, B_{i},\theta_1)|} +
        \frac{f_{\theta_{{i \; {\rm
        new}}(\theta_2)}}}{|dg/d\theta(A_{i}, B_{i},\theta_2)|} +
        \cdots + \frac{f_{\theta_{{i \; {\rm new}}}(\theta_n)}}{|dg/d\theta(A_{i}, B_{i},\theta_n)|}.
\label{eq:fepsilon}
\end{equation}
The difficulty here is to find the roots $\theta_1, \theta_2,
\cdots,\theta_n$. The equation (\ref{eq:roots}) is
fourth order in $\theta$ (consequently the maximum value for $n$ is 4)
and there are analytical expressions to compute the roots, 
but they are too CPU-time demanding to be computed
for every site $i$.
Moreover, the exact calculation of these roots for other models 
could be impossible.

For these reasons we propose to calculate the roots numerically and to
store a table $f_{\varepsilon_{i \; {\rm new}}}(A_i, B_i,
\varepsilon)$ for some values of $A_i$, $B_i$ and $\varepsilon$. Other
values not stored in the table can be founded by interpolation
techniques.  This approach is very fast, since we need only to calculate 
$A_i$, $B_i$ and $\varepsilon_i$ for each selected spin and to read
from the table the value for the respective p.d.f..  Figure (1) shows
the p.d.f. $f_{\varepsilon_i}(\varepsilon_i)$ for some values of $A_i$
and $B_i$, obtained from a table with $128$ entries for $A_i$, $128$
for $B_i$ and $256$ for $\varepsilon$. Despite of these few values we
did not find any significant difference between using linear
interpolation or taking the nearest value in the table.  Furthermore,
nearly the same results are obtained with only $128$ entries for
$\varepsilon_i$.

Because $\Delta E$$=$$\varepsilon_{i \; {\rm new}}$$-$$\varepsilon_i$, 
its p.d.f. for a change in site $i$ is  
$f^i_{\Delta E}(\Delta E)$$=$$f_{\varepsilon_{i \; {\rm new}}}(A_i, B_i, \varepsilon_i+\Delta E)$.
Finally, since all sites are equally probable,  for a given configuration {\bf X} we obtain
\begin{equation} 
  f_{\Delta E}({\bf X}, \Delta E) = \frac{1}{N}\sum_{i}{f_{\varepsilon_{i \; {\rm new}}}(A_i, B_i,
  \epsilon_i+\Delta E)}.
\end{equation}
where the sum is over all sites.

\section{Sampling Procedure}
\label{sec:implement}

Toward a finite-size scaling analysis of this
two-dimensional model we investigated system sizes of
$L$$=$$6$,$8$,$10$,$16$,$20$,$40$,$60$ and $100$ with periodic boundary
conditions.  A single value of $\Delta E$$=$$4.0$ was chosen for the BHM
analysis.  Since we are interested only in thermodynamic
functions near the critical temperature it is not necessary to sample the
whole energy range of positive temperatures. Instead, the following
schema was adopted: For each system size we set a maximal and
a minimal energy values $E_{\rm max}$ and $E_{\rm min}$ and chose
$NE$ equally-spaced energy points between them.  Our goal is to
estimate $\beta (E)$ at each point $E$, as
\begin{eqnarray}
\beta (E) = \frac{1}{\Delta E_b} \ln \left(
         \frac{\langle N(E-\Delta E_b/2, \Delta E) \rangle}
         {\langle N(E+\Delta E_b/2, -\Delta E)\rangle} \right) + O(\Delta E_b^2) \; ,
\label{beta3_BHM}
\end{eqnarray}
and to integrate it by the trapezoidal rule.

Samples were taken by performing microcanonical simulations
on small windows of width $\Delta E_b$, that choose and turn one spin
at random and accept the new state only if it falls into the window.
This procedure takes all states inside the window with the same probability.
Then, most of the samples will cumulate on the high-energy border of the
window, because the density of states is a fast-growing function.
Therefore, $\langle N(E+\Delta E_b/2, -\Delta E)\rangle$ is well
estimated from a microcanonical simulation in a window of width
$\Delta E_b$ centered at $E$, and $\langle N(E-\Delta E_b/2, \Delta E)
\rangle$ from a microcanonical simulation in a window of width $\Delta
E_b$ centered at $E-\Delta E_b$.  A schematic description of this
sampling procedure is shown in figure (\ref{XYZ2Microsim}).  Table
\ref{TableXYZ2sims} shows the values of $E_{\rm max}$ $E_{\rm min}$
and $NE$ used for each system size.

Summarizing, we ran two microcanonical simulations for each value of
$\beta (E)$.  The microcanonical averages $\langle M \rangle_E$,
$\langle M^2 \rangle_E$ and $\langle M^4 \rangle_E$ were computed at
each energy point by averaging the $M$-values in both windows.  Inside
each window $1000$ MCSS were discarded before sampling and, then,
$100$ samples spaced by $10$ MCSS were taken.  The whole procedure was
repeated $20$ times and the data of all $20$ runs were cumulated in
the same histograms.

\section{Finite-Size Scaling Analysis}
\label{sec:results}

We performed a finite-size scaling analysis \cite{barber74,binder81}
to estimate the critical properties of this model.
We have divided our analysis in two parts; first, the determination of
the critical temperature and, second, the estimation of the critical
parameters.

\subsection{Critical temperature and Binder cumulant}

One of the questions we want to elucidate is the dependence of the
results on the random number generator employed. Hence, we repeated
our simulations with four different random number generators, i.e.,
{\textsc RAN2} \cite{press92}, {\textsc RANLUX} with the fourth {\it
  luxury} level \cite{james94}, {\textsc RANIBM} \cite{press92} and a
lagged-{\textsc FIBONACCI} \cite{knuth81,marsaglia85} random number
generator with primitive trinomial (110503,53719,0) \cite{heringa92}, 
initialized with {\textsc RAN2}.

Figure (\ref{Ubinder}) shows the fourth-order Binder cumulant
\begin{eqnarray}
  U_L(T)=1-{1 \over 3}{\langle M^4 \rangle_T \over \langle M^2 \rangle_T^2} 
\label{EqUBinder}
\end{eqnarray}
computed for all system sizes with the four random number generators.
By looking at the limits of the smallest rhombus crossed by all curves
we define maximal and minimal values for the critical coupling
$\beta_{T_{\rm c}}$ and the universal parameter $U^*:=U_L(T_{\rm c})$
(table \ref{TableXYZ2_UTc}).  The most precise results are obtained
with the {\textsc FIBONACCI} and the less ones with {\textsc RANLUX}.
By combining the four results through a maximal likehood estimator
\cite{brownlee65} we obtain $T_{\rm c}=1.5329(4)$ ($\beta_{\rm
  c}=1/k_{\rm B}T_{\rm c}=0.6524(2)$) and $U^*=0.6118(6)$ , in good
agreement with the value $U^*=0.618$ for the 2D-Ising universality
class \cite{binder85} (here and in the following the number between
parenthesis represents a $2\sigma$ error bar, with a confidence level
of $95$\%.  For instance, $2.158(24)=2.158 \pm 0.024$ and $\sigma=0.012$).

To corroborate this value for the critical temperature we have plotted
the temperatures of  maximal heat capacity $T_{Cv}(L)$ and maximal
magnetic susceptibility  $T_{Xs}(L)$ as a function of  the system size
$L$.
In the thermodynamic limit 
\begin{eqnarray}
        T_{Cv}(L)=T_{\rm c}+ a L^{-1/\nu}  \mbox{\hspace{3mm};\hspace{3mm}} 
        T_{Xs}(L)=T_{\rm c}+ b L^{-1/\nu} \; .
\label{TCvXs}
\end{eqnarray}
Since $\nu=1.0$ for the 2D-Ising universality class, we expect a
linear behavior of $T_{Cv}(L)$ and $T_{Xs}(L)$ against $1/L$. Figure
(\ref{FigureXYZ2_TCvTXs}) shows this linear behavior with the four
random number generators. The linear regressions were obtained from
the data points at $L=10,16,20,40,60$ and $100$.  All figures coincide
to estimate a critical temperature near $T_{\rm c}=1.53$, in excellent
agreement with our estimate from the Binder cumulant and in
concordance with the value of $T_{\rm c}=1.529(5)$ obtained in ref.
\cite{resende98} with {\textsc RAN2}, but far away of value $T_{\rm
  c}=1.491(6)$ obtained in the same reference with {\textsc RANLUX}.

\subsection{Critical exponents}

To compute the critical exponents of this model we used the 
data obtained with the {\textsc FIBONACCI} random number generator,
that has showed to be the most precise one.

The critical exponent $\nu$ can be obtained by two different procedures.
First, it
can be estimated from the previous data for the Binder cumulant
by using the relation \cite{binder81}
\begin{eqnarray}
   \left. {dU_L \over dT} \right|_{T_{\rm c}}
   \sim - L^{-1/\nu} \; .
\label{nu}
\end{eqnarray}
Figure (\ref{FigureXYZ2nu}a) shows the power-law fit for this
relation at $T_{\rm c}=1.5329$.  The value $1/\nu=1.009(18)$
($\nu=0.991(16)$) obtained is consistent with the value $\nu=1.0$ of
the 2d-Ising universality class.  Second, it can be estimated from
(\ref{TCvXs}) by plotting $T_{Xs}(L)-T_{\rm c}$ against $1/L$ (figure
(\ref{FigureXYZ2nu}b)).  This second value of $1/\nu=0.920(24)$
($\nu=1.087(28)$) is not so precise as the first one.  By combining
both results in a maximal likehood estimator, we obtain
$\nu=1.015(16)$.

Figure (\ref{FigureXYZ2beta}a) shows the average magnetization per site
$\langle M \rangle_{T_{\rm c}}$ at $T_{\rm c}=1.5329$ 
as a function of the system size $L$,
in order to compute the
critical parameter $\beta / \nu$ from the finite-size behavior 
\begin{eqnarray}
  {1 \over L^d} \langle M \rangle_{T_{\rm c}} \sim L^{-\beta / \nu} \; .
\end{eqnarray}
The value of $\beta / \nu=0.1340(34)$ is higher than 
the value $1.25$ of the 2d-Ising universality class.
This can be a small-sizes effect, since the same estimation excluding $L=6,8,10$
and $16$ gives $\beta / \nu=0.1285(20)$.
By excluding in consecutively the system sizes  $L=6,8,10$ and $16$ we obtained
different values of $\beta/ \nu $. These values,  when plotted against $1/L$, 
show a linear behavior ($L$ is the minimal system size taken into account)
(figure (\ref{FigureXYZ2beta}b)).
The linear extrapolation gives a final estimation for $\beta / \nu=0.126(2)$.
Figure (\ref{FigureXYZ2beta}c) shows the FSS collapse of the average magnetization,
by taking $\nu=1.015$ and $\beta / \nu=0.126$.

To compute $\gamma / \nu$ we have used the three finite-size scaling relations
\cite{binder69,paauw75,bernreuther88,binder81}
\begin{eqnarray}
  {1 \over L^d} Xs(T_{\rm c}) &\sim& L^{\gamma / \nu} \quad , \\
  \label{EqFSSXs1}
  {1 \over L^d} Xs_{\rm max}  &\sim& L^{\gamma / \nu} \quad , \\
  \label{EqFSSXs2}
  {1 \over L^d} \langle M^2 \rangle_{T_{\rm c}} &\sim& L^{\gamma / \nu} \quad .
  \label{EqFSSXs3}
\end{eqnarray}
Figure (\ref{FigureXYZ2gamma}a) shows the results form these three
methods.  The results are $\gamma / \nu=1.745(8)$ (eq.
\ref{EqFSSXs1}), $\gamma / \nu=1.741(4)$ (eq. \ref{EqFSSXs2}) and
$\gamma / \nu=1.798(4)$ (eq. \ref{EqFSSXs3}).  By excluding
consecutively the system sizes $L=6,8,10$ and $16$ we obtain the
values shown in figure (\ref{FigureXYZ2gamma}b).  The larger variation
corresponds to the estimations from $Xs_{\rm max}$, as expected
\cite{peczak91,ferrenberg91a}.  The linear extrapolation of these
values does not coincide for $L\to\infty$, Nevertheless, they all
intersect around $\gamma / \nu = 1.742(4)$, when the system sizes
$L=6,8$ and $10$ have been excluded. Furthermore, it coincides with
the maximal-likehood estimator of the three values, $\gamma /
\nu=1.742(4)$, in good agreement with the value for the 2d-Ising
universality class, $\gamma / \nu=1.75$.  Figure
(\ref{FigureXYZ2gamma}c) shows the FSS collapse of the magnetic susceptibility
with $\nu=1.015$ and $\gamma / \nu=1.742$.

The more difficult parameter to be directly computed is $\alpha$, the critical parameter
for the divergence of the specific heat.
On one side, assuming hyperscaling $\alpha=2-d\nu$ to be valid and setting
our estimate of $\nu=1.015(16)$ we obtain for  $d=2$ (the lattice dimension)
$\alpha=-0.030(32)$, which corresponds to a finite cusp,
\begin{eqnarray}
        C_{v_{\rm max}}(L) = C_{v_{\rm max}}(L=\infty) - aL^{\alpha / \nu} \quad .
\label{XYZ2_Cvpeak}
\end{eqnarray}
On the other side, for the 2d-Ising universality class $\alpha=0.0$ and 
$Cv_{\rm max}$ shows a logarithmic divergence \cite{fisher67}
\begin{eqnarray}
        Cv_{_{\rm max}}(L) = a \ln{L} + b + O[\ln{L}/L] \quad .
\label{XYZ2_Cvlogmax}
\end{eqnarray}
Furthermore, one predicts the scale behavior \cite{ferdinand69,barber74}
\begin{eqnarray}
        C_{v_L}(x) = A_0 \ln{L} + Q^{(0)}(x) + O[\ln{L}/L] \quad ,
\label{XYZ2_CvlogT}
\end{eqnarray}
where $x \sim L(T-T_{\rm c})/T_{\rm c}$.
The constant $A_0$ can be deduced from the scaling behavior of $C_{v_L}(T_{\rm c})$
\begin{eqnarray}
        C_{v_L}(T_{\rm c})=A_0 \ln{L} + Q^{(0)}(0) + O[\ln{L}/L] \quad .
\label{XYZ2_CvlogTc}
\end{eqnarray}

Figure (\ref{FigureXYZ2alpha}a) shows both a power-law fit with
$\alpha / \nu=-0.039$ ($C_{v_{\rm
    max}}(L)=25.0+3.200(4) \times L^{-0.03938(14)}$) and a logarithmic fit
($C_{v_{\rm max}}(L)=0.62(10)+0.853(31) \ln{L}$) for $C_{v_{\rm
    max}}(L)$ (circles).  The value of $C_{v_{\rm max}}(\infty)=25.0$ for
the first one was set by looking for the larger correlation
coefficient $r$ of the power-law fit.  This maximal value
$r=0.9989758$ changes, however, only in the fifth decimal place for
$16.0<C_{v_{\rm max}}(\infty)<297.0$, corresponding to
$-0.0732(28)<\alpha<0.00281(10)$.  We take, hence, $\alpha=-0.04(8)$,
in good agreement with the value from hyperscaling.  The logarithmic
fit, with correlation coefficient $r=0.99888879$ is almost
indistinguishable from the power-law fit.

Figure (\ref{FigureXYZ2alpha}a) also shows a logarithmic fit for
$C_{v_L}(T_{\rm c})$ (squares). This fit estimates $A_0=0.818(20)$
from (\ref{XYZ2_CvlogTc}) with correlation coefficient $r=0.9995723$.
Figure (\ref{FigureXYZ2alpha}b) shows the FSS collapse of the specific heat
according with (\ref{XYZ2_CvlogT}) for this value of $A_0$. All system
sizes collapse well, except for $L=100$ at $T<T_{\rm c}$, due,
perhaps, to a lack of simulational points for this system size.
All these results support the
thesis of a logarithmic anomaly of the specific heat for this model.

Our results are summarized in table \ref{TableXYZ2CriticalParameters}.
They are in good agreement with the expected values for the
2d-Ising universality class and \cite{resende98}.

\section{Comments and Conclusions}
\label{sec:conclusion}

We have constructed a BHM analysis procedure for the XY-model with
$Z_2$ symmetry. With this example, we have shown how to construct a
BHM analysis when the quantities $N(E,\Delta E)$ cannot be easy
computed.  Our strategy is to construct a table for the function
$f^i_{\varepsilon_i}$ (eq. \ref{eq:fepsilon}), where the indexes are
given by the variables which define the energy per spin
$\varepsilon_i$, and the values of the function are determined by
numerical methods.  Other values not stored in the table are found by
interpolation techniques.  This approach is very fast, since only a
memory access, an interpolation and a sum are needed to compute
$N(E,\Delta E)$. It is only restricted by the amount of memory
available.  This approach is general and can be extended to other
complex models.

We have employed the BHM analysis procedure together with microcanonical
simulations to perform a finite-size scaling analysis of the 
XY-model with $Z_2$ symmetry, for system sizes $6 \le L \le 100$. 
Our calculations give a very precise estimates for the critical temperature,
$T_{\rm c}=1.5329(4)$ and of the Binder cumulant $U^*=0.6118(6)$. 
Furthermore, our results for $T_{\rm c}$ and $U^*$ 
are almost the same for the four 
pseudo-random number generators employed, {\textsc RAN2}, {\textsc RANLUX},
{\textsc FIBONACCI} and {\textsc RANIBM}, with the most precise results for
{\textsc FIBONACCI} and the least ones for {\textsc RANLUX}.
Our estimations for the critical exponents are in good agreement within the error bars
with the theoretical predictions of the 2d-Ising universality class and with
other Monte Carlo studies.

Our value of $T_{\rm c}$ agrees with the result $T_{\rm c}=1.529(5)$
from \cite{resende98} with {\textsc RAN2} and clearly disagrees with the result
$T_{\rm c}=1.491(6)$ from the same reference with {\textsc RANLUX}.
To clarify this point, figure (\ref{NewCosta}) shows our results for the
magnetic susceptibility in a $20\times20$ system, obtained with
different random number generators and including also some reference
points calculated with the Metropolis algorithm, which reproduce the
figure (1) of ref. \cite{resende98}.  In that previous work, the
Metropolis algorithm with {\textsc RANLUX} showed a peak at a clearly
lower temperature ($T=1.55$) than the same algorithm with {\textsc
  RAN2} ($T=1.58$).  In contrast, all our data coincide to show a peak
around $T=1.58$, including the Metropolis estimations with {\textsc
  RANLUX}.  In our opinion, the result of $T_{\rm c}=1.491(6)$ could
be due to an incorrect implementation of {\textsc RANLUX}.

The computational time of the BHM analysis increases proportional to
$N$, the number of sites, in the same proportion as the time of one
MCSS with the microcanonical sampling procedure. A special care should
be taken, however, to assure that enough microcanonical simulations
contribute to each temperature average.  According to eq. (\ref{meanQE}), 
the contribution of a sample with energy $E$ to
the average $\langle Q \rangle_T$ is weighted by the factor $g(E)
\exp(-E(x)/k_{\rm B}T)$, which distributes approximately as a Gaussian
with center at $\langle E \rangle_T$ and width $\sigma_E(T)$$=$$T
\sqrt{k_{\rm B}C_v(T)}$.  Therefore, all samples in the range $\langle
E \rangle_T-3\sigma_E<E<\langle E \rangle_T+3\sigma_E$ will contribute
to $\langle Q \rangle_T$.  For small system sizes, the whole range of
energies contributes.  For large ones, we have tried to maintain a
constant number of simulations inside $1 \sigma_E$.  The number of
simulations per energy interval can be chosen by assuming $C_v(T)$$\sim$$
L^d$ for $\alpha\le0$ and $C_v(T) \sim L^{d+\alpha / \nu}$ otherwise.
Table \ref{TableXYZ2_NEperSigma} shows the number of microcanonical
simulations in $1 \sigma_E$ at $x=L(T-T_{\rm c})/T_{\rm c}=-1$ by
assuming $C_v(x=-1)=0.818 \ln{L} - 0.5$ (see figure
(\ref{FigureXYZ2alpha}b)).

In conclusion, 
we have shown that the BHM method can be used to obtain 
quality results from the finite-size scaling analysis of a non-trivial system,
also if the quantities required to implement the method are hard to compute
analitically.
These results support the usefulness and generality of the BHM.

\section*{Acknowledgments}
We want to thank very specially Hans J. Herrmann 
for encouraging discussions and suggestions in the develop of this work.
ARL wishes to thank H.J. Herrmann and J. D. Mu\~noz for the hospitality and
financial support at ICA1 where this work was done.  The authors acknowledge
financial support from CNPq (Brazilian Agency) (ARL) and Deutscher
Akademischer Austauschdienst (DAAD) (JDM, through scholarship A/96/0890).

\newpage

\begin{table}[hb]
  \begin{center}
    \begin{tabular}{|c||c|c|c|c|c|c|c|c|} 
      L   &   6 &   8 &  10 &  16 &   20 & 40 & 60 & 100 \\ 
      \hline
      $E_{\rm max}$ & 0.0 & 0.0 & 0.0 & 0.0 &-0.35 &-0.35 &-0.35 &
      -0.35 \\ 
      \hline
      $E_{\rm min}$ &-1.0 &-1.0 &-1.0 &-1.0 &-0.70 &-0.70 &-0.70 &
      -0.70 \\ 
      \hline
      $NE$          & 200 & 200 & 200 & 320 & 140  & 280  & 420  & 700 \\ 
    \end{tabular}
  \end{center}
  \caption[]{Parameters of the microcanonical simulations.}
  \label{TableXYZ2sims}
\end{table}

\begin{table}[b]
  \begin{center}
    \begin{tabular}{|c||c|c|c|c|c|} 
      Generator   & RAN2 & RANLUX & FIBONACCI & RANIBM & Combined   \\ 
      \hline
      $T_{\rm c}$ & 1.5317(16) & 1.5306(27) & 1.5330(4) & 1.5328(13) &
      1.5329(4) \\ 
      \hline
      $U^*$       & 0.6110(29) & 0.6139(32) & 0.6118(6) & 0.6109(42) & 0.6118(6) \\ 
    \end{tabular}
    \end{center}
  \caption[]{Critical temperature $T_{\rm c}$ and Binder cumulant $U^*$
    for the two-dimensional XY-model with $Z_2$ symmetry.}
\label{TableXYZ2_UTc}
\end{table}

\begin{table}[h]
  \begin{center}
    \begin{tabular}{|c||c|r|r|r|r|}
      & Generator         &   $\nu$  & $\beta/\nu$ & $\gamma/\nu$ & $\alpha/\nu$  \\ \hline\hline
      This study               &{\textsc FIBONACCI} & 1.015(16) &   0.126(2)  &    1.742(8)  &   -0.04(8)    \\ \hline
      Ref. \cite{resende98} &{\textsc RANLUX}    &    -      &   0.14(6)   &    1.71(7)   &    0.00(3)    \\ \hline
      Ref. \cite{resende98} &{\textsc RAN2}      &    -      &   0.13(4)   &    1.75(9)   &    0.05(4)    \\ \hline
      2d Ising                 &                    & 1.000(0)  &   0.125(0)  &    1.750(0)  &    0.00(0)    \\ 
    \end{tabular}
  \end{center}
\caption[]{Critical Parameters for the XY-model with $Z_2$ symmetry.}
\label{TableXYZ2CriticalParameters}
\end{table}

\begin{table}[hb]
  \begin{center}
    \begin{tabular}{|c||r|r|r|r|r|r|r|r|}
      L &  6 &  8 & 10 & 16 & 20 & 40 & 60 & 100 \\ \hline
      $NE/\sigma_E$ & 14 & 12 & 11 & 13 & 14 & 16 & 17 & 18 \\ 
    \end{tabular}
  \end{center}
\caption[]{Number of microcanonical simulations at 
$x=-1$ for the settings of Table \ref{TableXYZ2sims}}
\label{TableXYZ2_NEperSigma}
\end{table}

\newpage
\begin{figure}[!h]
  \caption{Probability distribution for : $A=0, B=4$ (dot-dashed line),
    $A=2.25, B=2.25$ (continuous line), $A=-1.75, B=2.25$ (dashed line) and
    $A=0.5625, B=-0.1875$ (long-dashed line)}
  \label{fig:pdf}
\end{figure}

\begin{figure}
  \caption{Sampling procedure.}
  \label{XYZ2Microsim}
\end{figure}

\begin{figure}
  \caption{Binder cumulant for the XY-model with $Z_2$ symmetry obtained
    with ({\bf a}) the {\textsc RAN2}, ({\bf b}) the {\textsc
      RANLUX}, ({\bf c}) the {\textsc FIBONACCI} and ({\bf d}) the
    {\textsc RANIBM} random number generators.}
  \label{Ubinder}
\end{figure}

\begin{figure}
  \caption{Size dependence of the location of the peak for the magnetic 
    susceptibility (circles) and the heat capacity (squares) obtained
    with ({\bf a}) the {\textsc RAN2}, ({\bf b}) the {\textsc RANLUX},
    ({\bf c}) the {\textsc FIBONACCI} and ({\bf d}) the {\textsc
      RANIBM} random number generators.}
  \label{FigureXYZ2_TCvTXs}
\end{figure}

\begin{figure}
\caption{
  Estimations of $\nu$ from the slope of the Binder cumulant $U_L$ at
  $T_{\rm c}$ ({\bf a}) and the position of the peak in the magnetic
  susceptibility $T_{Xs}$ ({\bf b}).  We adopted $T_{\rm c}=1.5329$.}
  \label{FigureXYZ2nu}
\end{figure}

\begin{figure}
  \caption{
    ({\bf a}) Estimations of $\beta / \nu$ from the average
    magnetization per site at $T_{\rm c}=1.5329$.  ({\bf b}) Linear
    extrapolation of this estimations. ({\bf c}) FSS of the average
    magnetization with $\nu=1.015$ and $\beta / \nu=0.126$ for system
    sizes $L=16,20,40,60$ and $100$.}
  \label{FigureXYZ2beta}
\end{figure}

\begin{figure}
  \caption{({\bf a}) Estimations of $\gamma / \nu$ from 
    the magnetic susceptibility per site at $T_{\rm c}=1.5329$
    (circles), the maximal value of magnetic susceptibility (squares)
    and the average value of the square of the magnetization at
    $T_{\rm c}$ (diamonds).  ({\bf b}) Linear extrapolation of this
    estimations. ({\bf c}) FSS of the magnetic susceptibility with
    $\nu=1.015$ and $\gamma=1.742$ for system sizes $L=16,20,40,60$
    and $100$.}
  \label{FigureXYZ2gamma}
\end{figure}

\begin{figure}
\caption{
  ({\bf a}) Power-law fit (dashed line) and logarithmic fit (solid
  line) of the maximal specific heat (circles) as a function of the
  system size $L$, and logarithmic fit (dot-dashed line) of the
  specific heat at $T_{\rm c}=1.5329$ (squares).  ({\bf b}) Finite
  size scaling of the specific heat assuming a logarithmic anomaly for
  system sizes $L=16,20,40,60$ and $100$.}
   \label{FigureXYZ2alpha}    
\end{figure}

\begin{figure}
\caption{
  Magnetic susceptibility for a $20$$\times$$20$ system, computed from
  BHM on pairs of microcanonical simulations (BHM-IN) and from
  Metropolis simulations (MC), both using several pseudo-random number
  generators.}
 \label{NewCosta}
\end{figure}

 \newpage \pagestyle{empty}
\begin{figure}
  \centerline{\epsfig{file=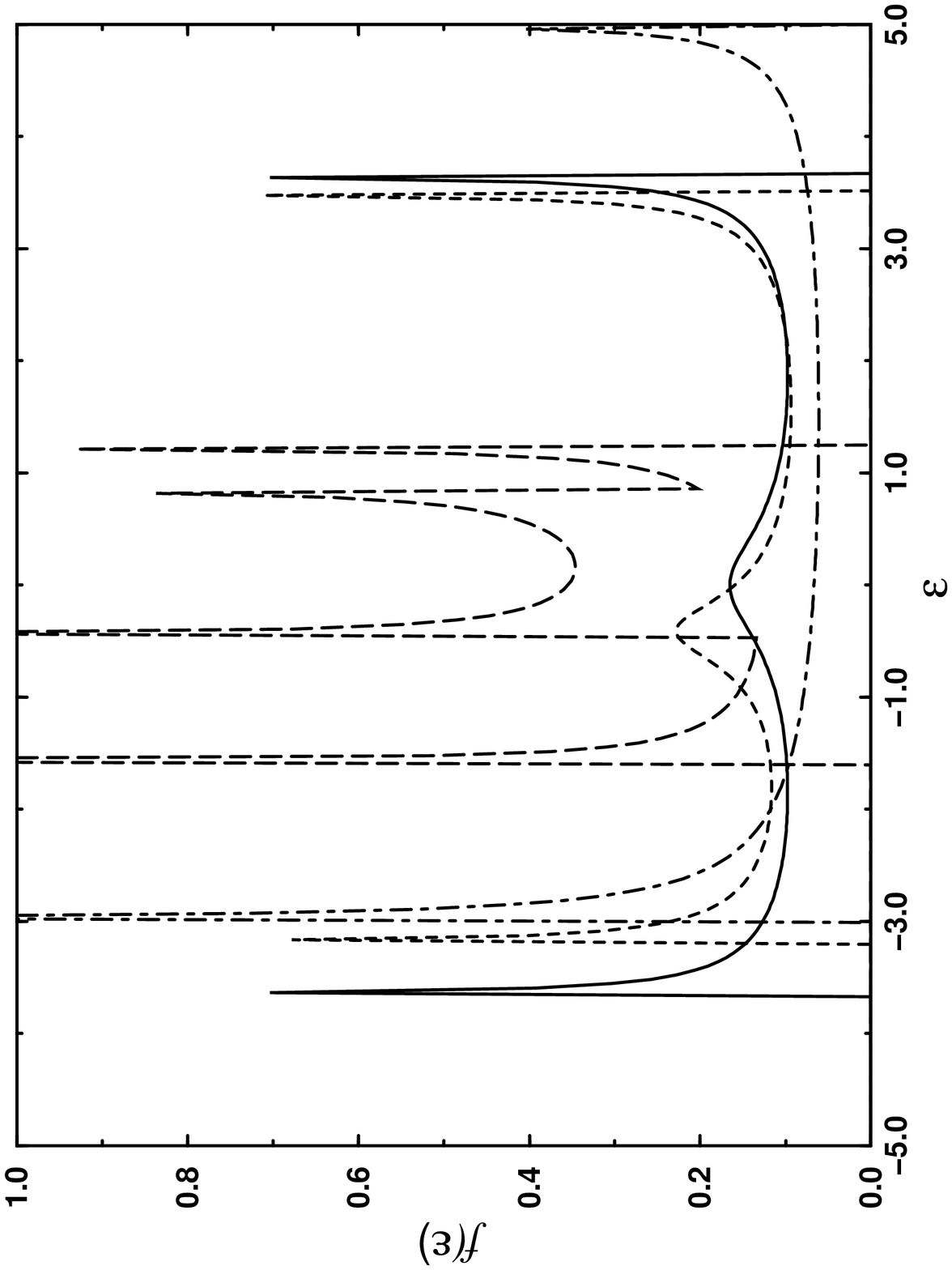,width=15cm}}
\end{figure}
\vfill
\centerline{Mu\~noz et. al., Figure 1}

\newpage
\pagestyle{empty}
\begin{figure}
  \centerline{\epsfig{file=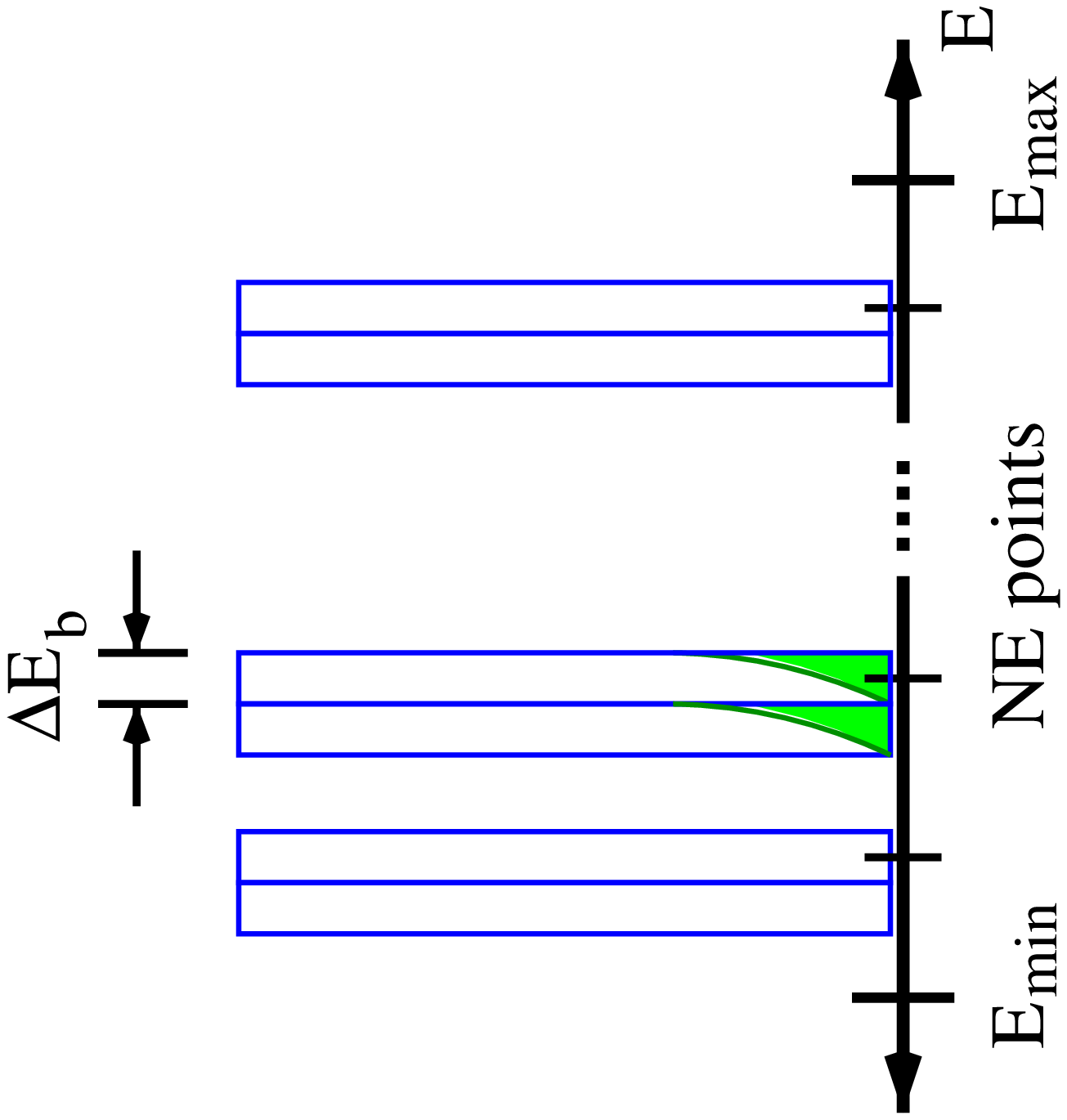,width=15cm,angle=90}}
\end{figure}
\vfill
\centerline{Mu\~noz et. al., Figure 2}

\newpage
\pagestyle{empty}
\begin{figure}
  \centerline{\epsfig{file=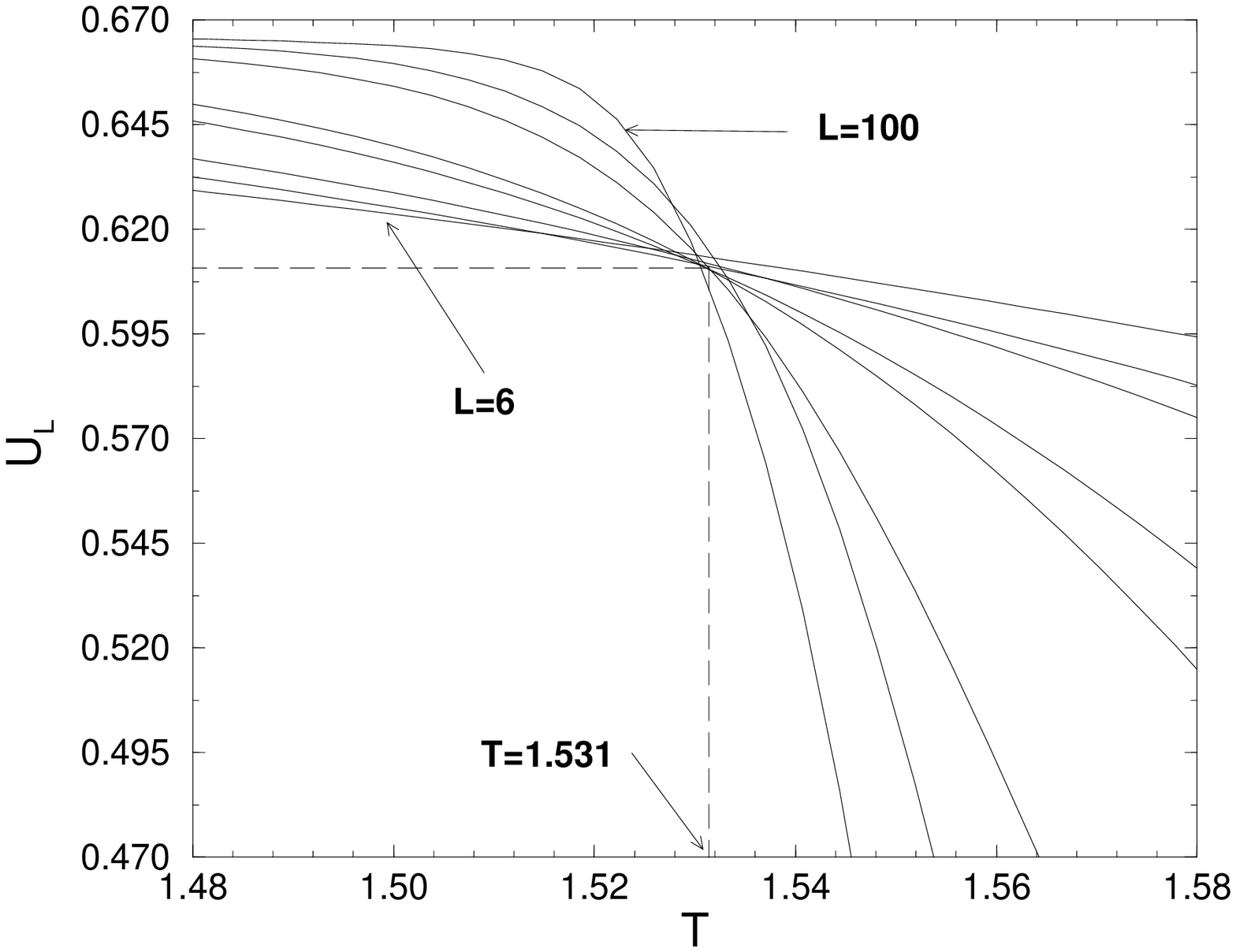,width=19cm,angle=90}}
\end{figure}
\vfill
\centerline{Mu\~noz et. al., Figure 3a}

\newpage
\pagestyle{empty}
\begin{figure}
  \centerline{\epsfig{file=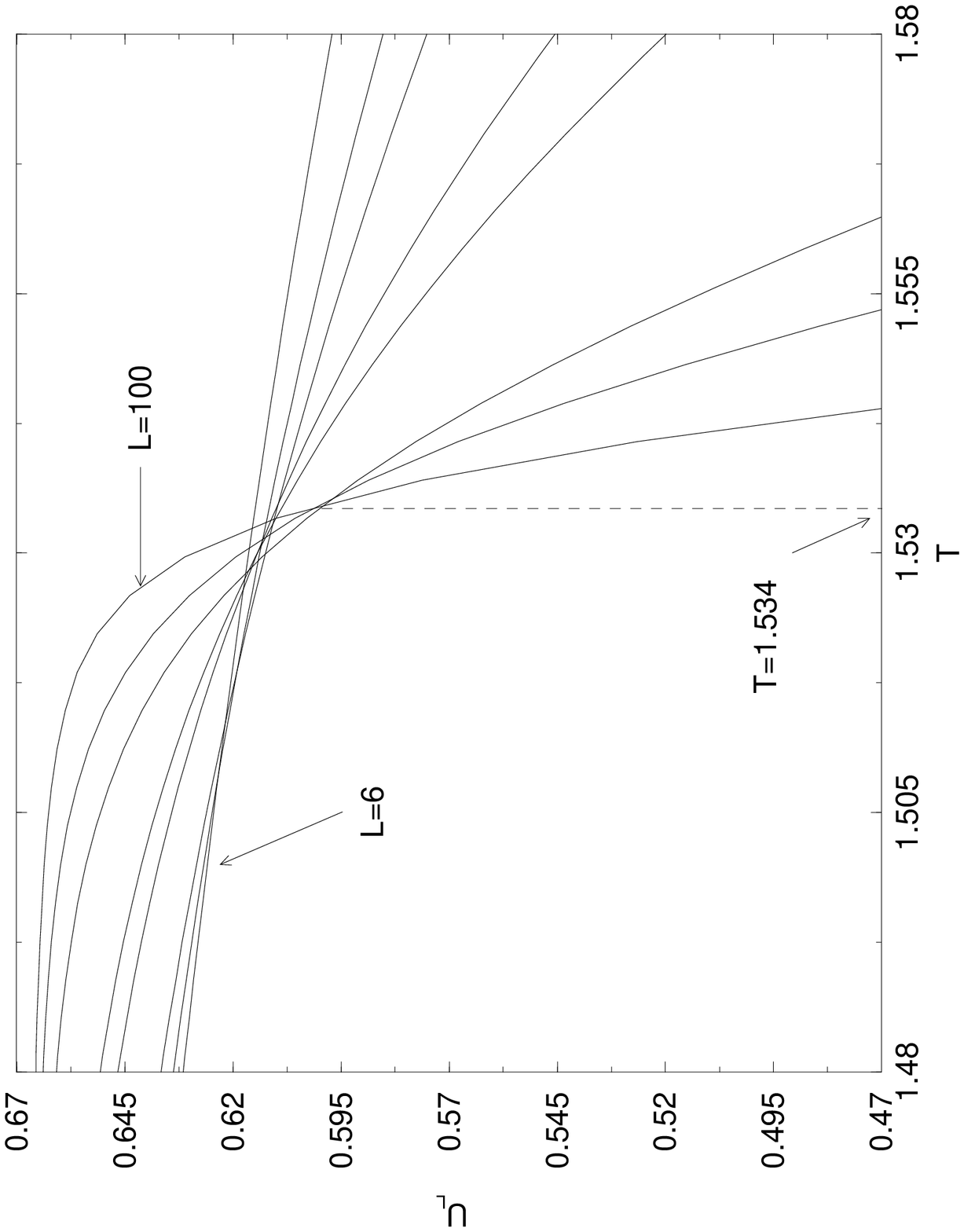,width=15cm}}
\end{figure}
\vfill
\centerline{Mu\~noz et. al., Figure 3b}

\newpage
\pagestyle{empty}
\begin{figure}
  \centerline{\epsfig{file=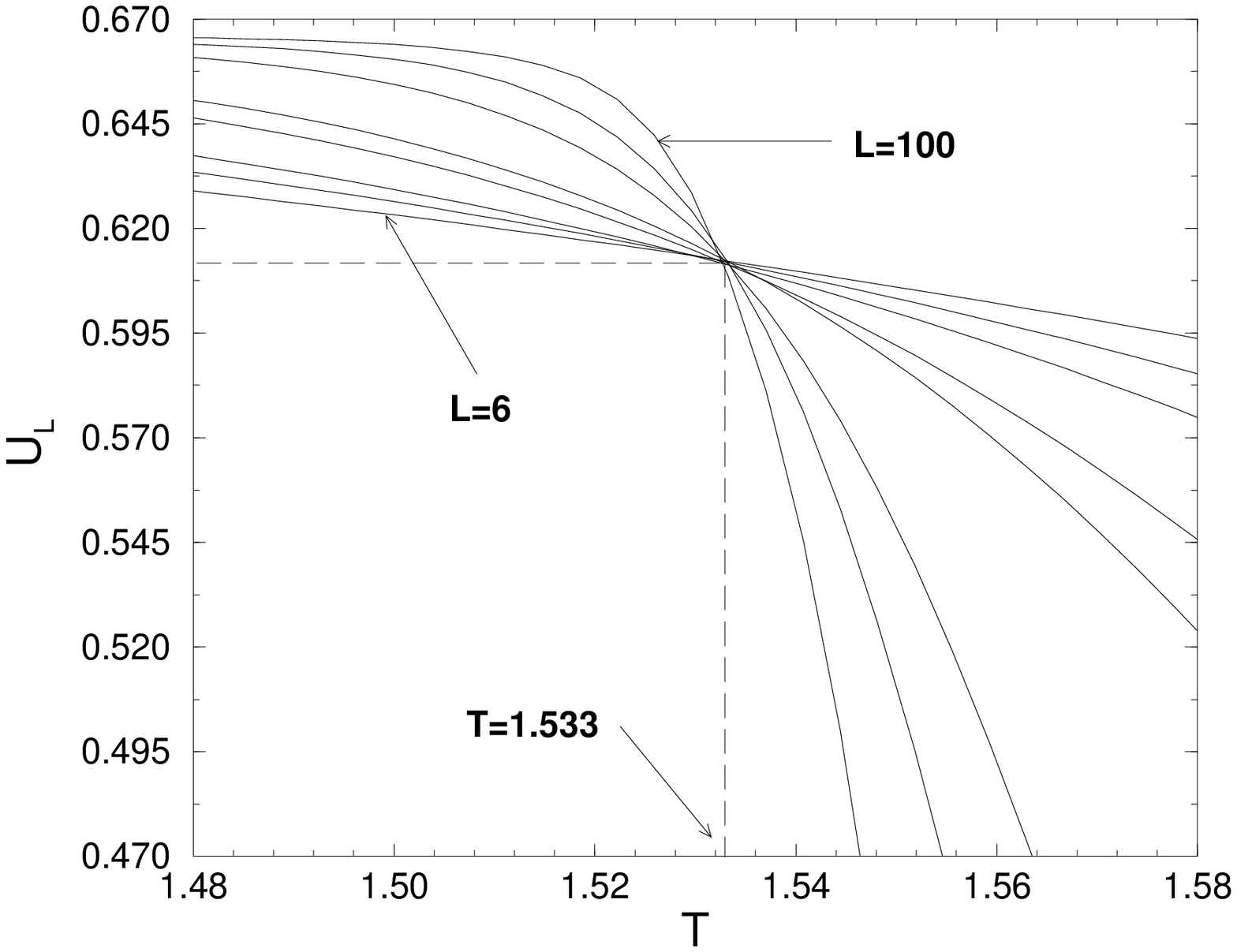,width=19cm,angle=90}}
\end{figure}
\vfill
\centerline{Mu\~noz et. al., Figure 3c}

\newpage
\pagestyle{empty}
\begin{figure}
  \centerline{\epsfig{file=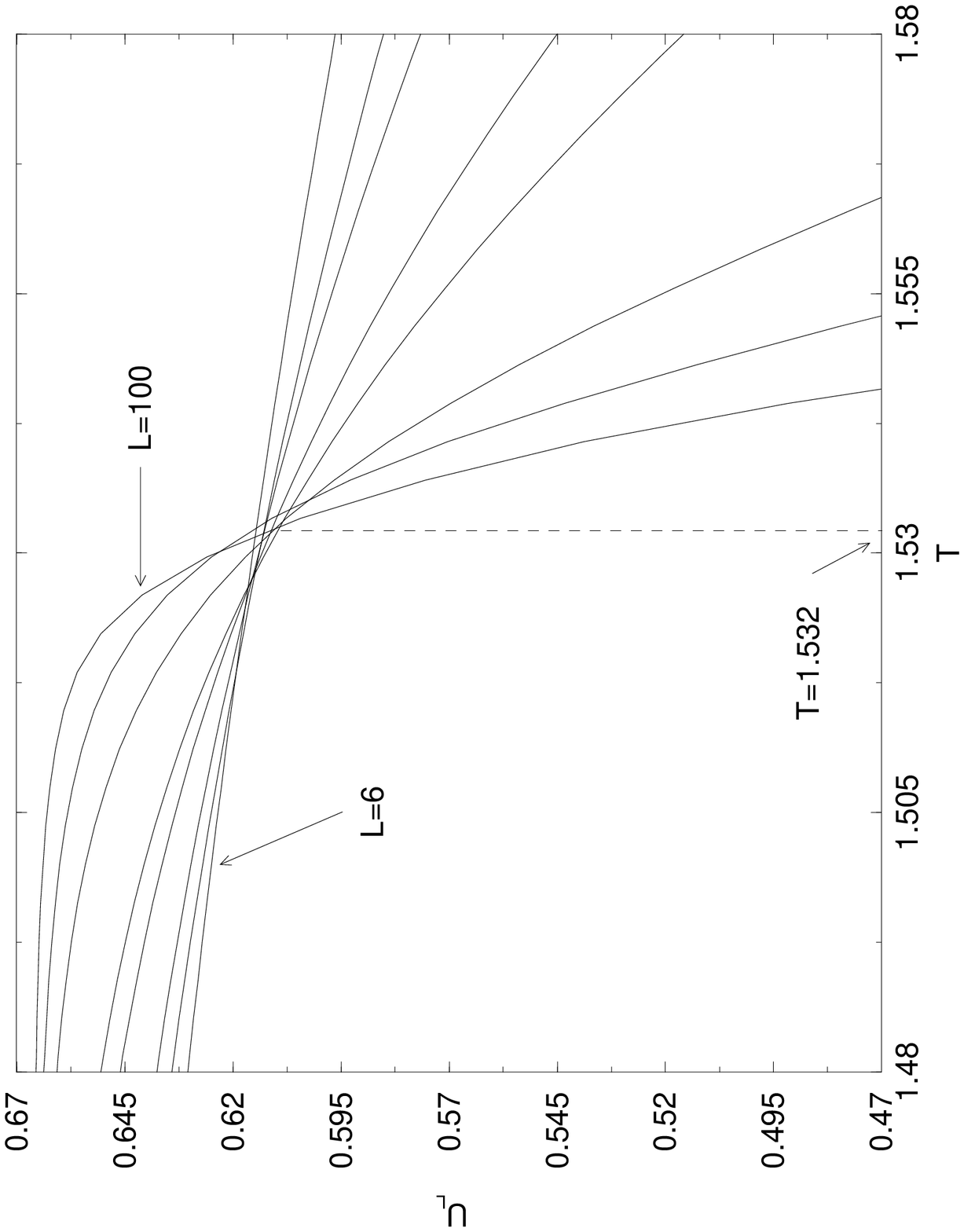,width=15cm}}
\end{figure}
\vfill
\centerline{Mu\~noz et. al., Figure 3d}

\newpage
\pagestyle{empty}
\begin{figure}
  \centerline{\epsfig{file=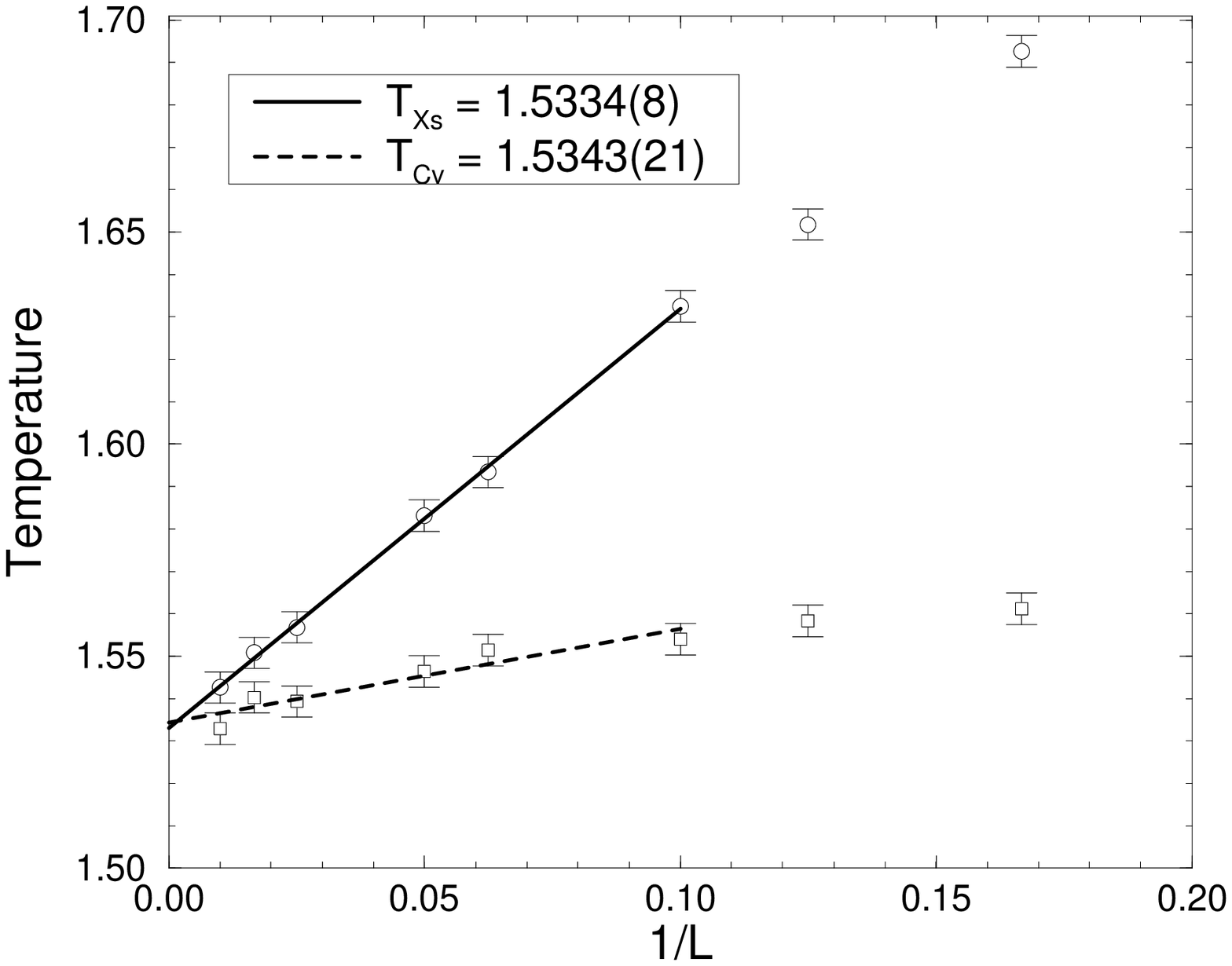,width=19cm,angle=90}}
\end{figure}
\vfill
\centerline{Mu\~noz et. al., Figure 4a}

\newpage
\pagestyle{empty}
\begin{figure}
  \centerline{\epsfig{file=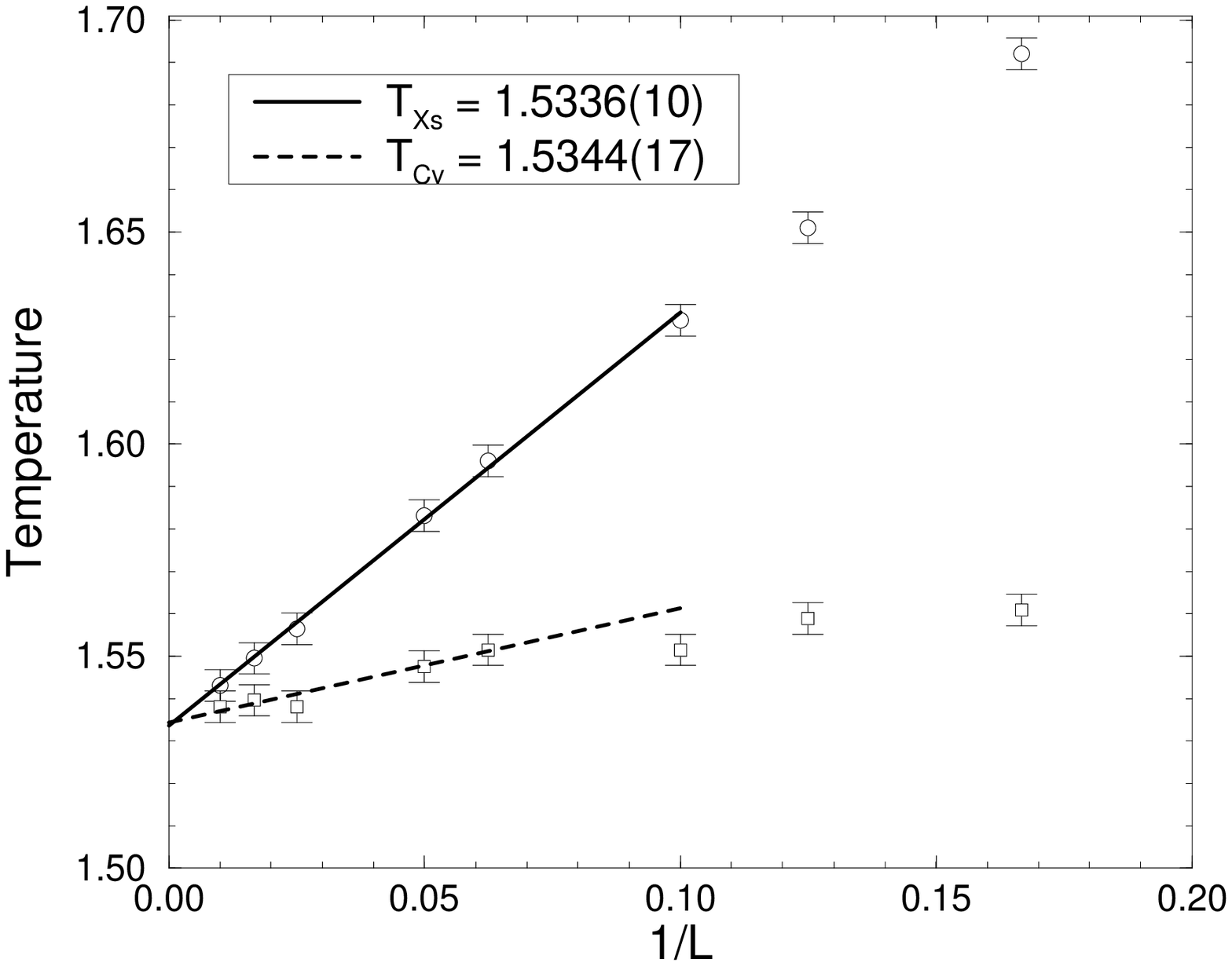,width=19cm,angle=90}}
\end{figure}
\vfill
\centerline{Mu\~noz et. al., Figure 4b}

\newpage
\pagestyle{empty}
\begin{figure}
  \centerline{\epsfig{file=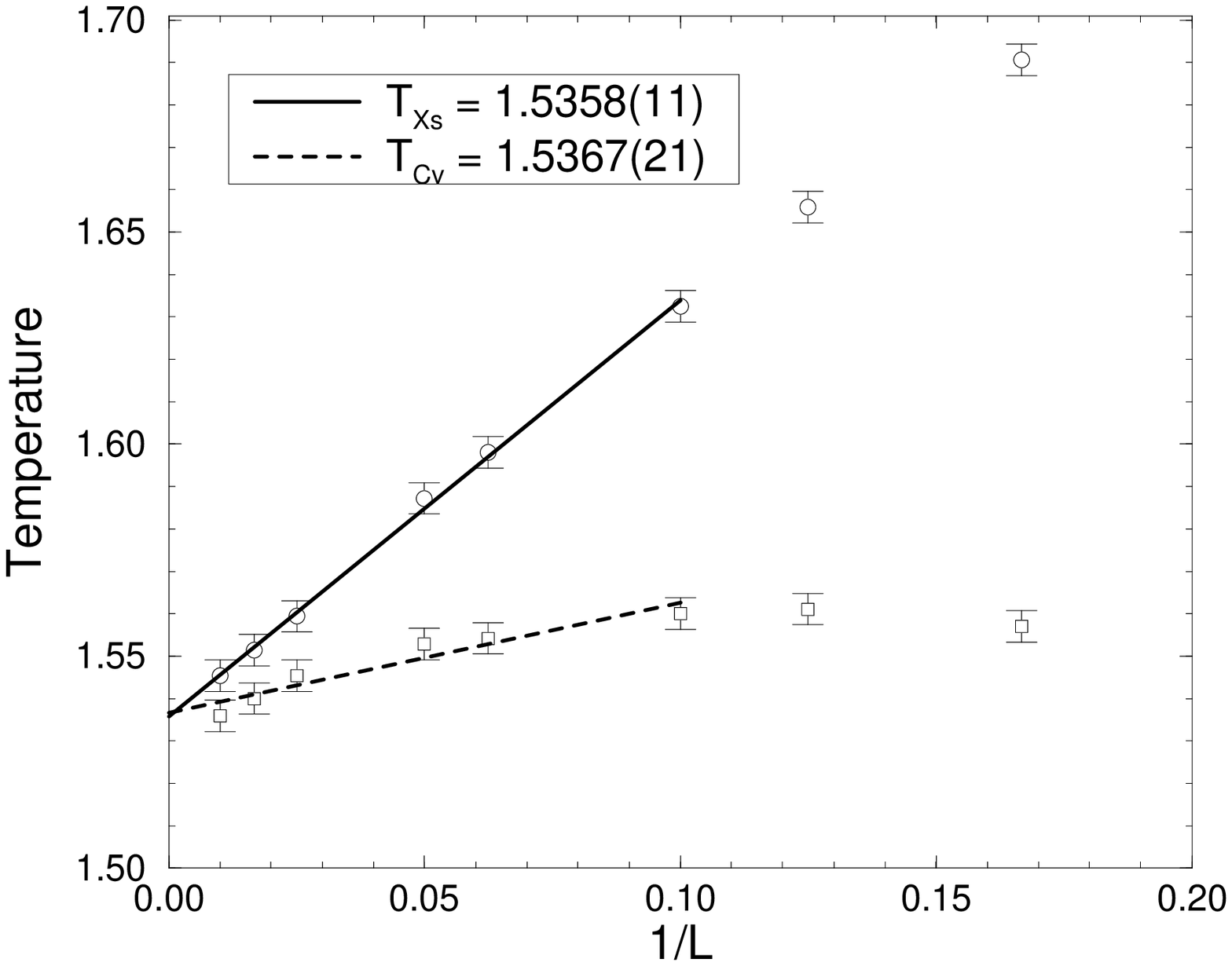,width=19cm,angle=90}}
\end{figure}
\vfill
\centerline{Mu\~noz et. al., Figure 4c}

\newpage
\pagestyle{empty}
\begin{figure}
  \centerline{\epsfig{file=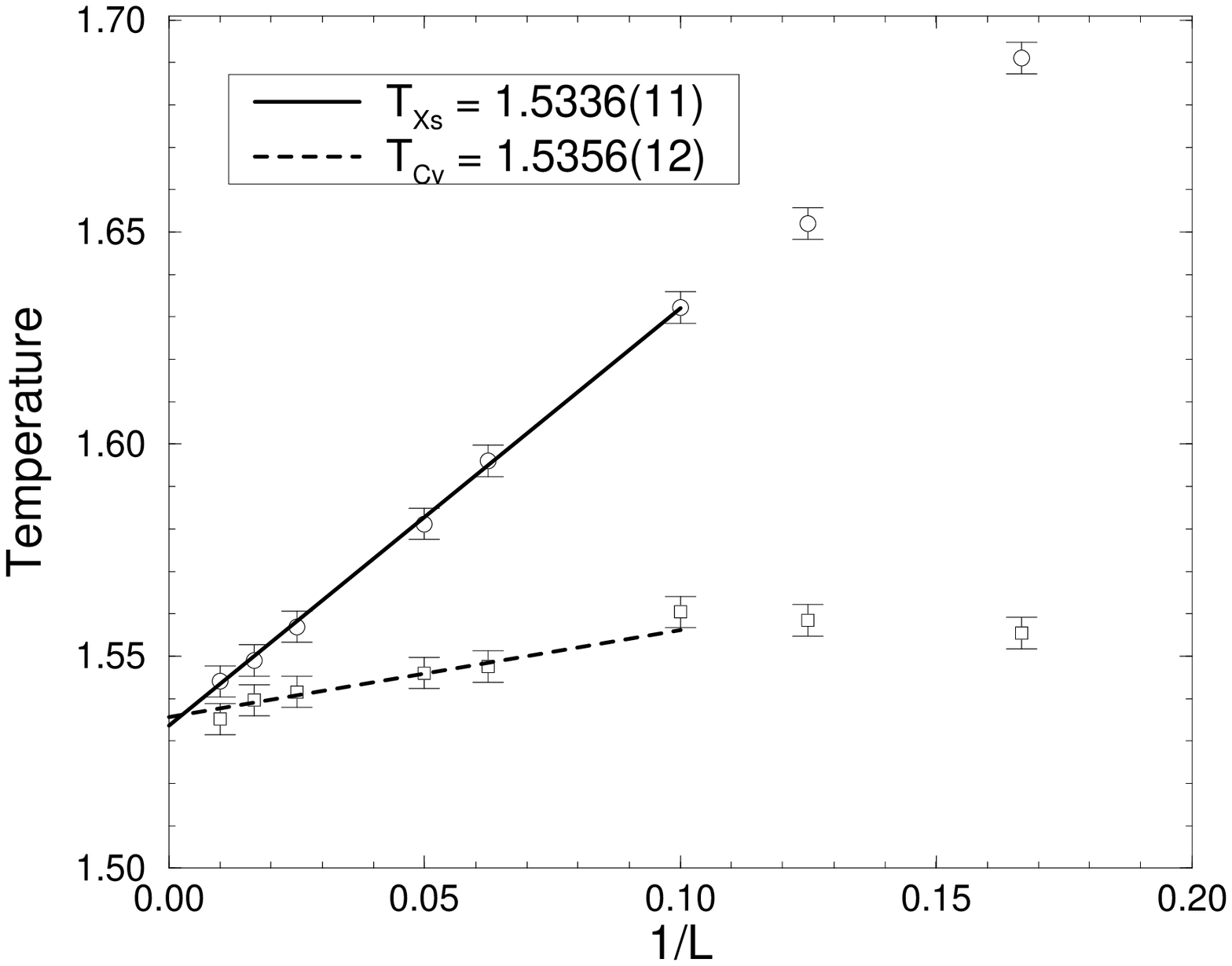,width=19cm,angle=90}}
\end{figure}
\vfill
\centerline{Mu\~noz et. al., Figure 4d}

\newpage
\pagestyle{empty}
\begin{figure}
  \centerline{\epsfig{file=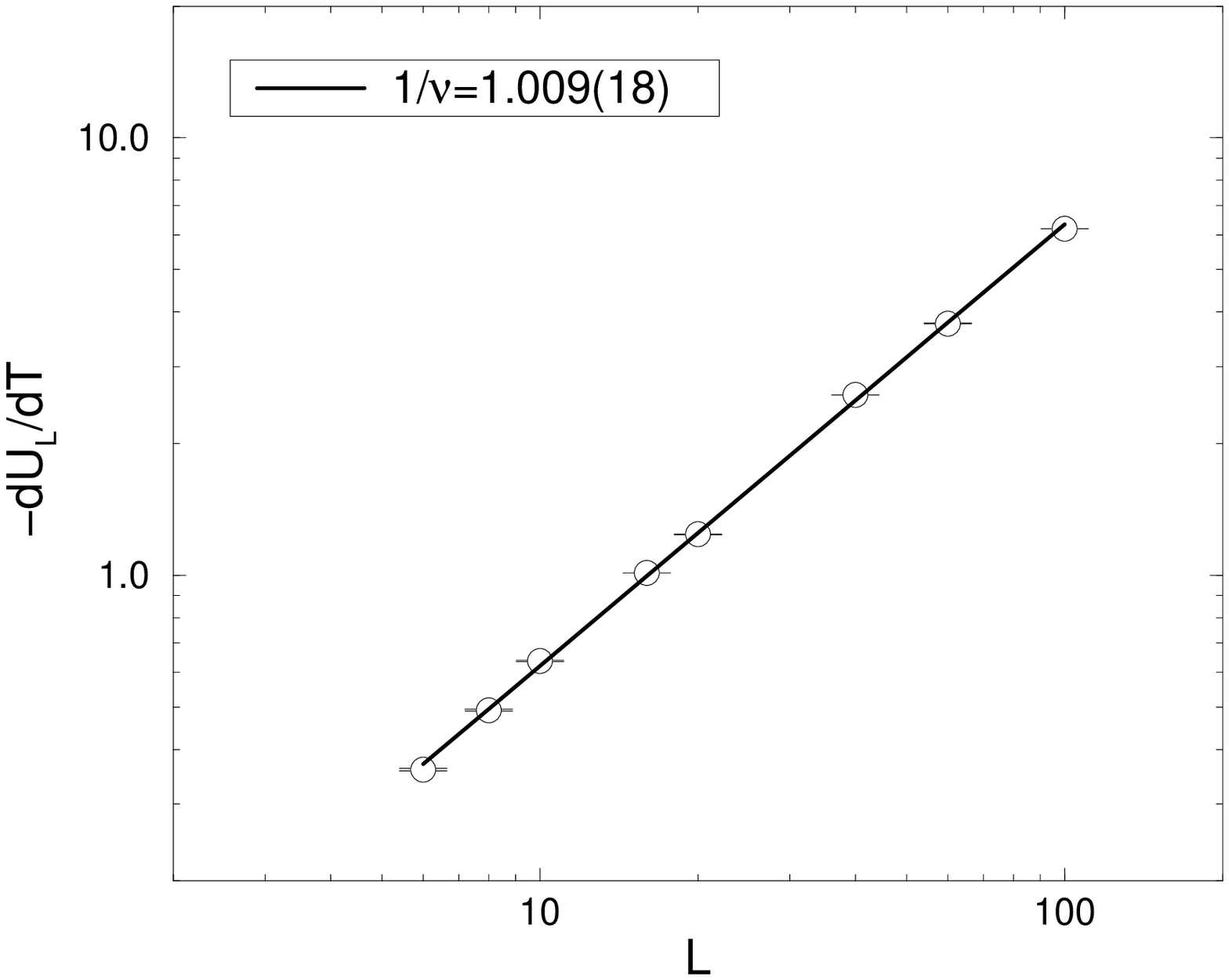,width=19cm,angle=90}}
\end{figure}
\vfill
\centerline{Mu\~noz et. al., Figure 5a}

\newpage
\pagestyle{empty}
\begin{figure}
  \centerline{\epsfig{file=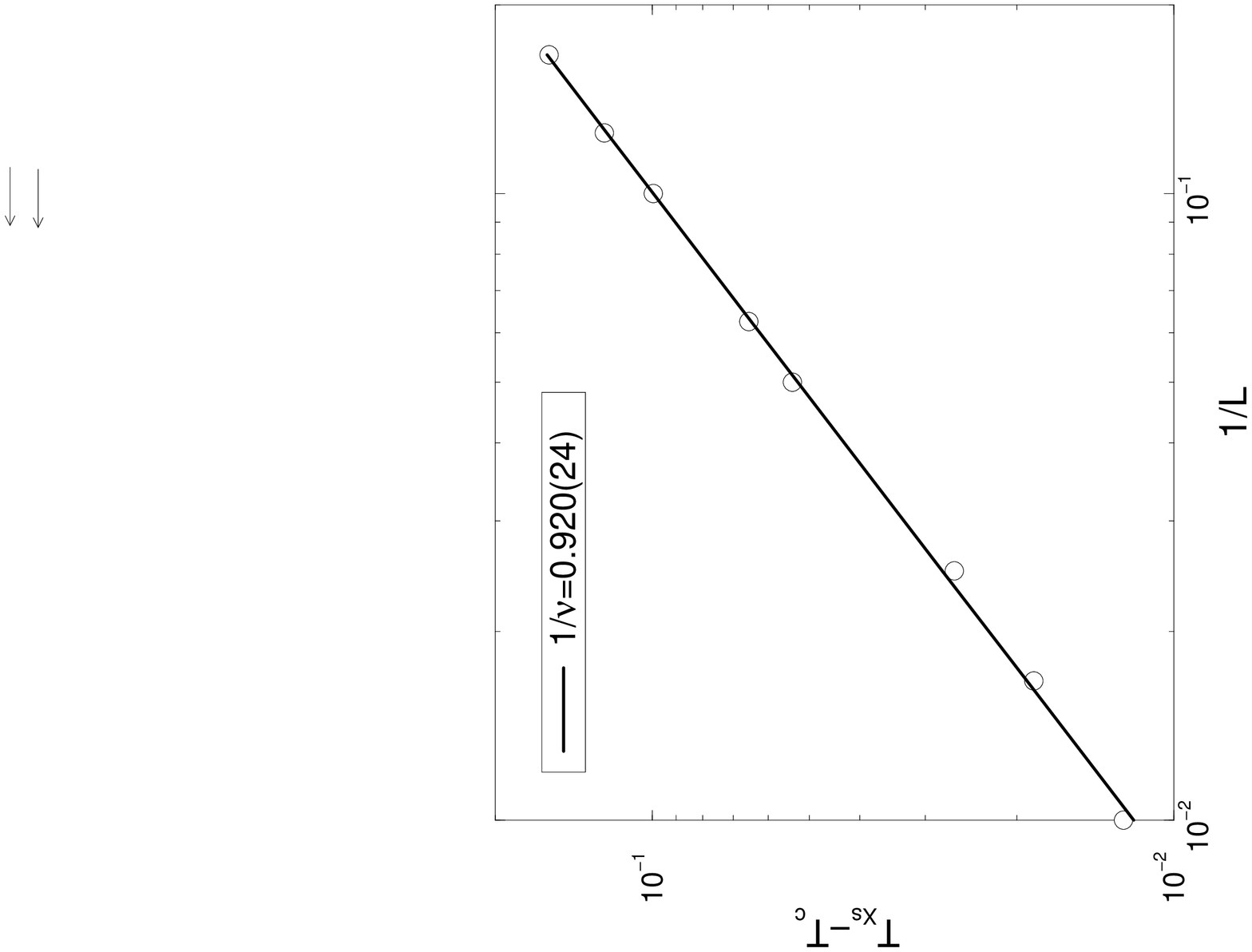,width=19cm,angle=90}}
\end{figure}
\vfill
\centerline{Mu\~noz et. al., Figure 5b}

\newpage
\pagestyle{empty}
\begin{figure}
  \centerline{\epsfig{file=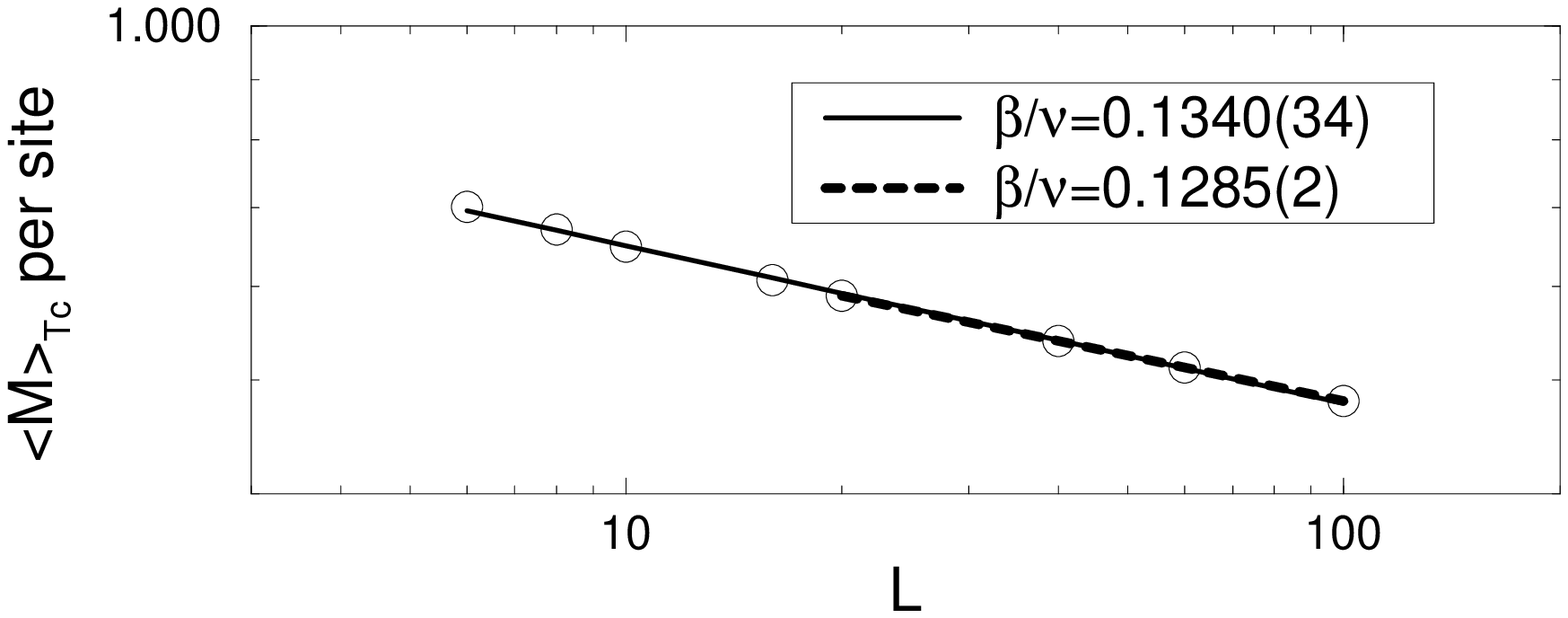,width=19cm,angle=90}}
\end{figure}
\vfill
\centerline{Mu\~noz et. al., Figure 6a}

\newpage
\pagestyle{empty}
\begin{figure}
  \centerline{\epsfig{file=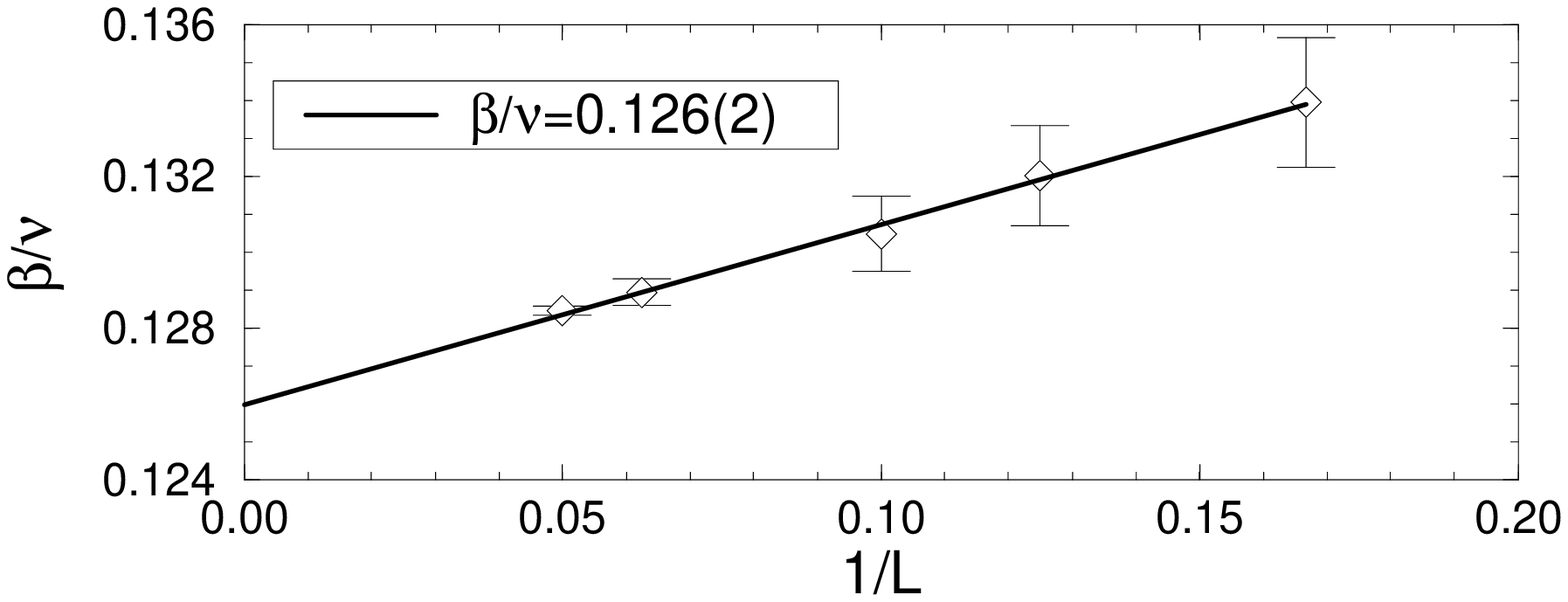,width=19cm,angle=90}}
\end{figure}
\vfill
\centerline{Mu\~noz et. al., Figure 6b}

\newpage
\pagestyle{empty}
\begin{figure}
  \centerline{\epsfig{file=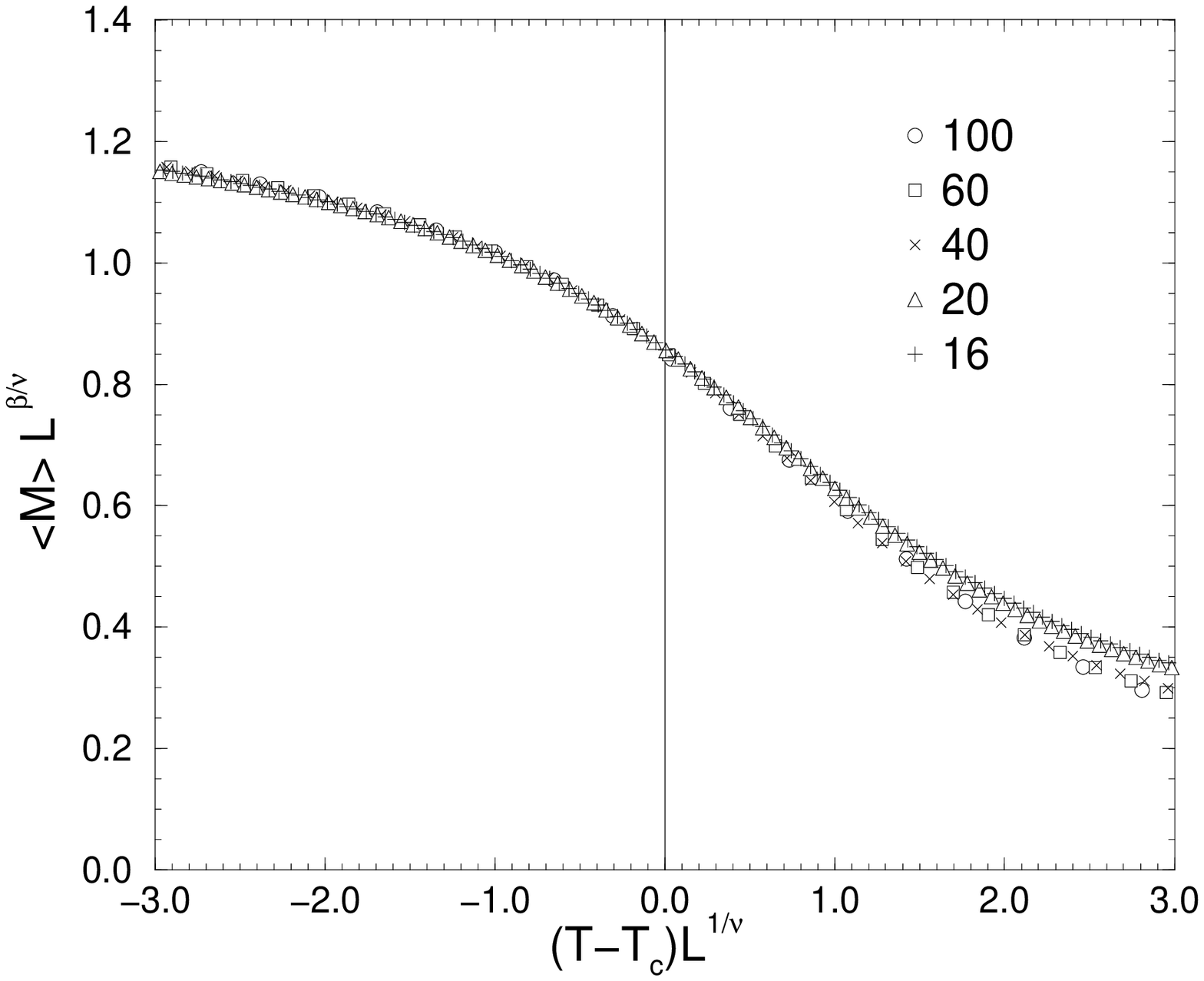,width=19cm,angle=90}}
\end{figure}
\vfill
\centerline{Mu\~noz et. al., Figure 6c}

\newpage
\pagestyle{empty}
\begin{figure}
  \centerline{\epsfig{file=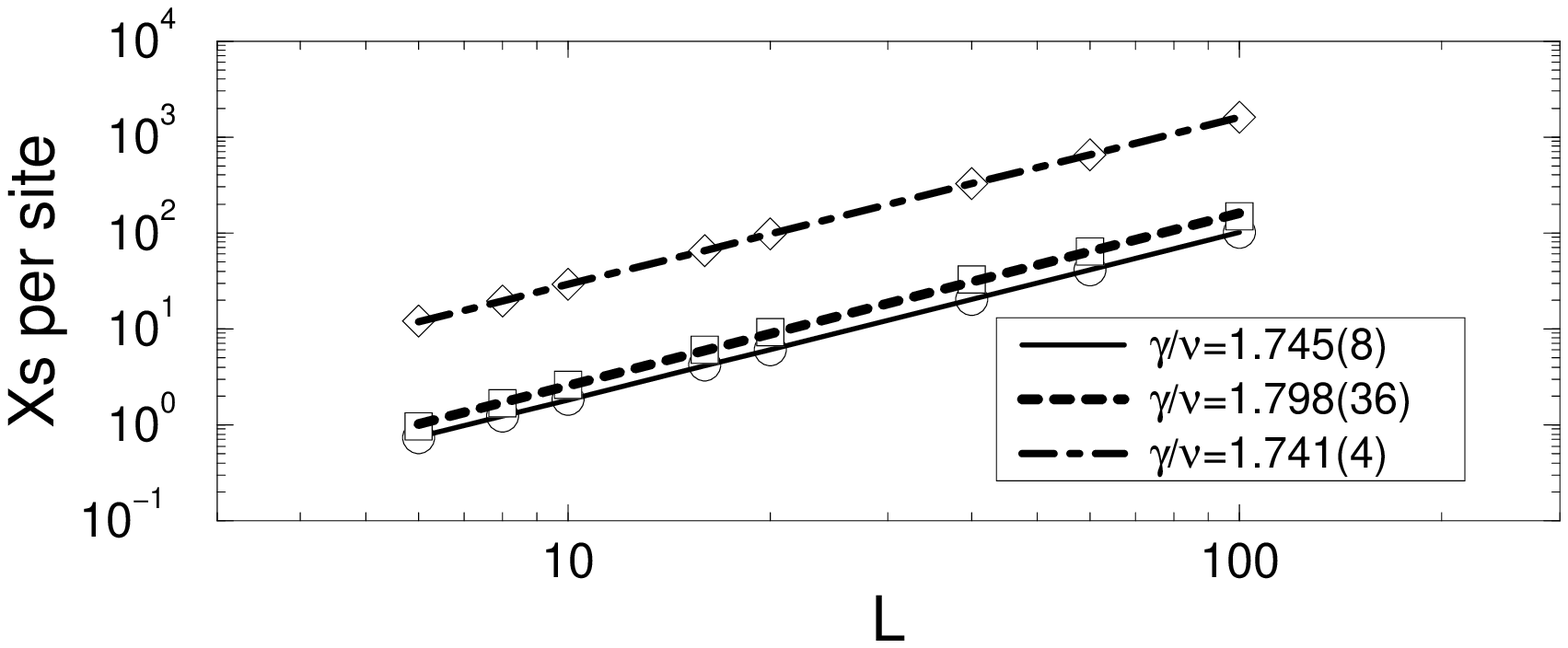,width=19cm,angle=90}}
\end{figure}
\vfill
\centerline{Mu\~noz et. al., Figure 7a}

\newpage
\pagestyle{empty}
\begin{figure}
  \centerline{\epsfig{file=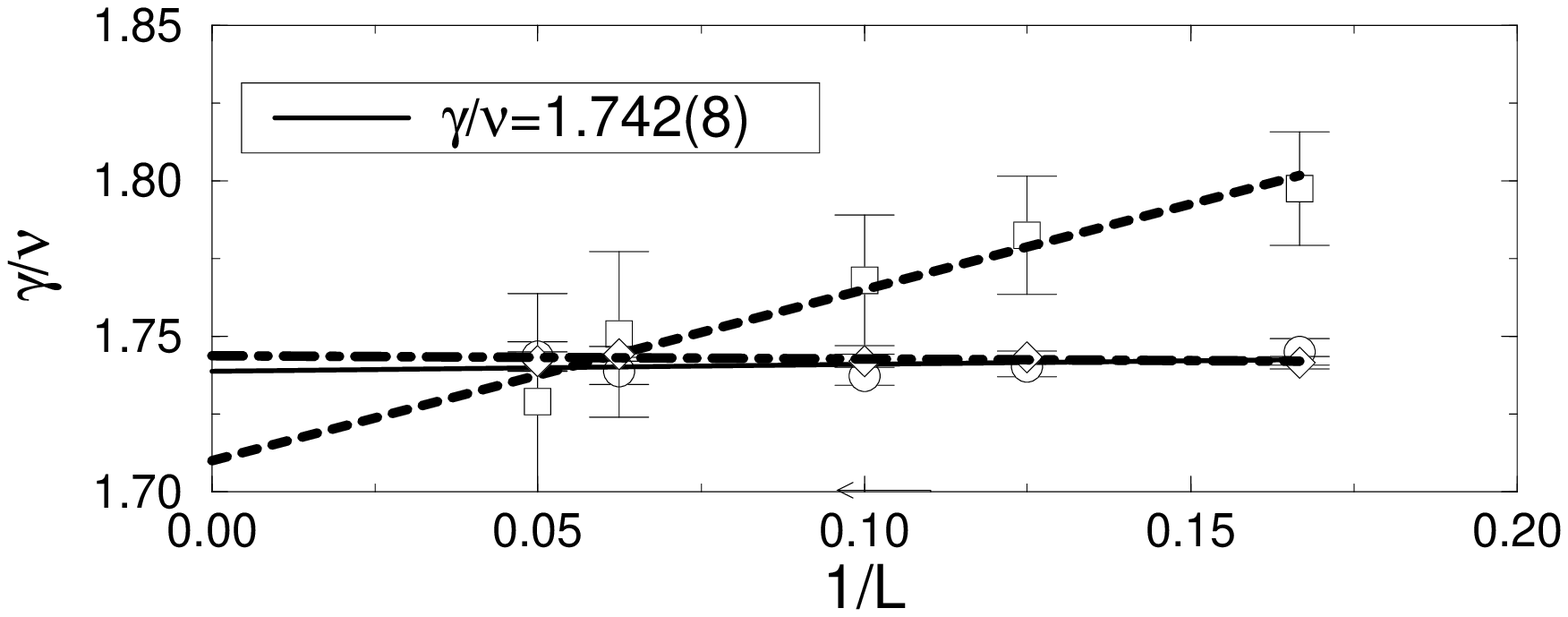,width=19cm,angle=90}}
\end{figure}
\vfill
\centerline{Mu\~noz et. al., Figure 7b}

\newpage
\pagestyle{empty}
\begin{figure}
  \centerline{\epsfig{file=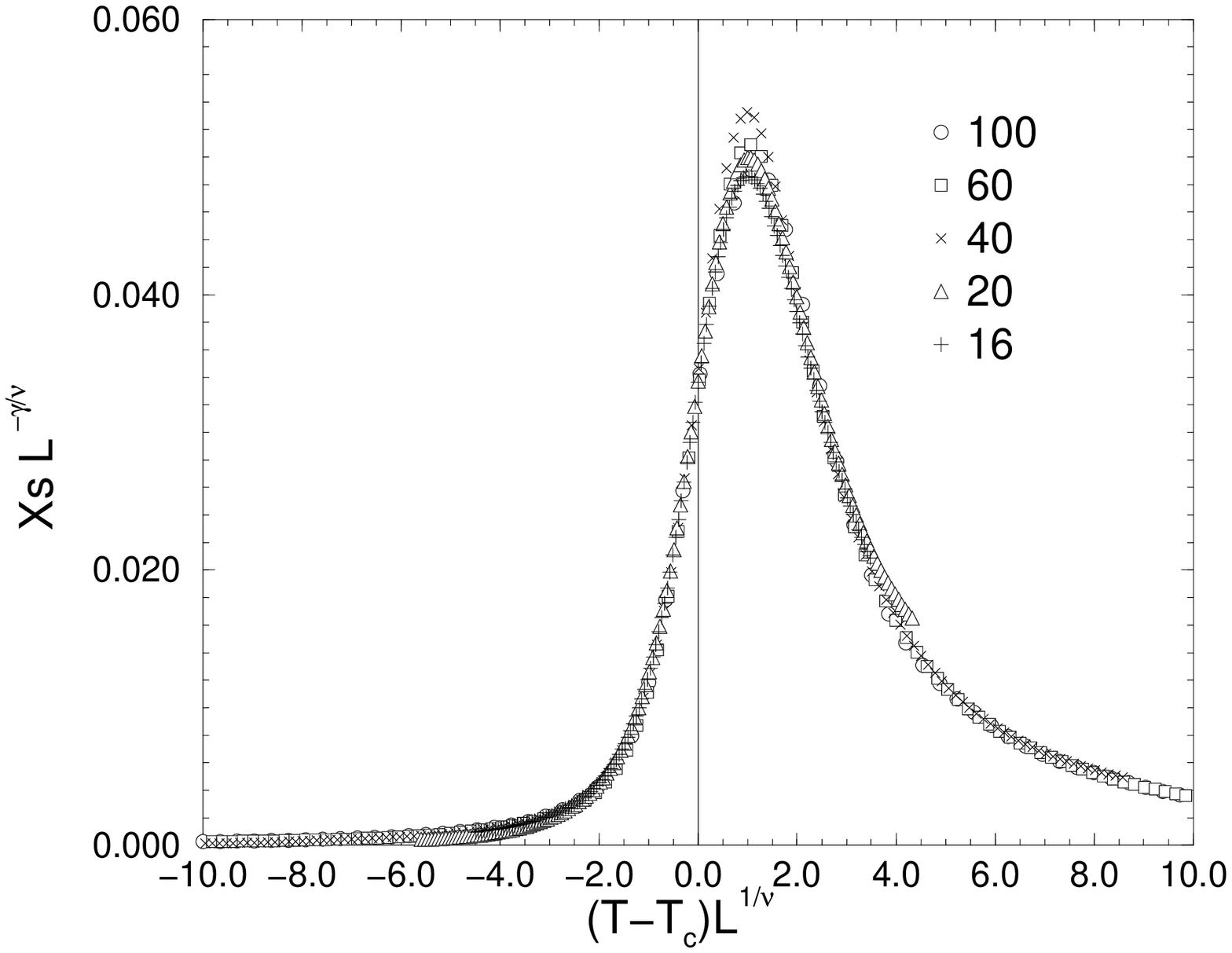,width=19cm,angle=90}}
\end{figure}
\vfill
\centerline{Mu\~noz et. al., Figure 7c}

\newpage
\pagestyle{empty}
\begin{figure}
  \centerline{\epsfig{file=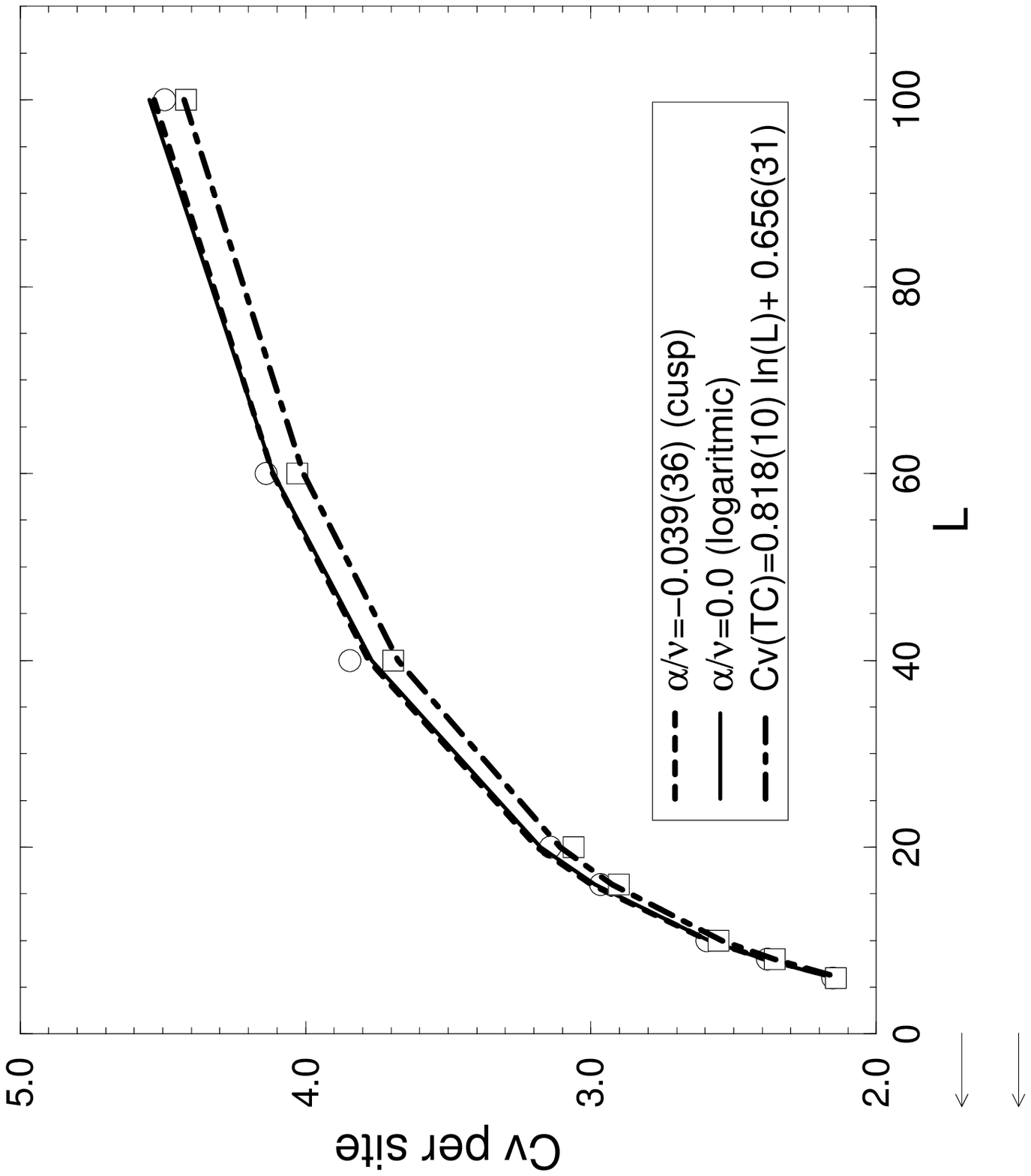,width=19cm,angle=90}}
\end{figure}
\vfill
\centerline{Mu\~noz et. al., Figure 8a}

\newpage
\pagestyle{empty}
\begin{figure}
  \centerline{\epsfig{file=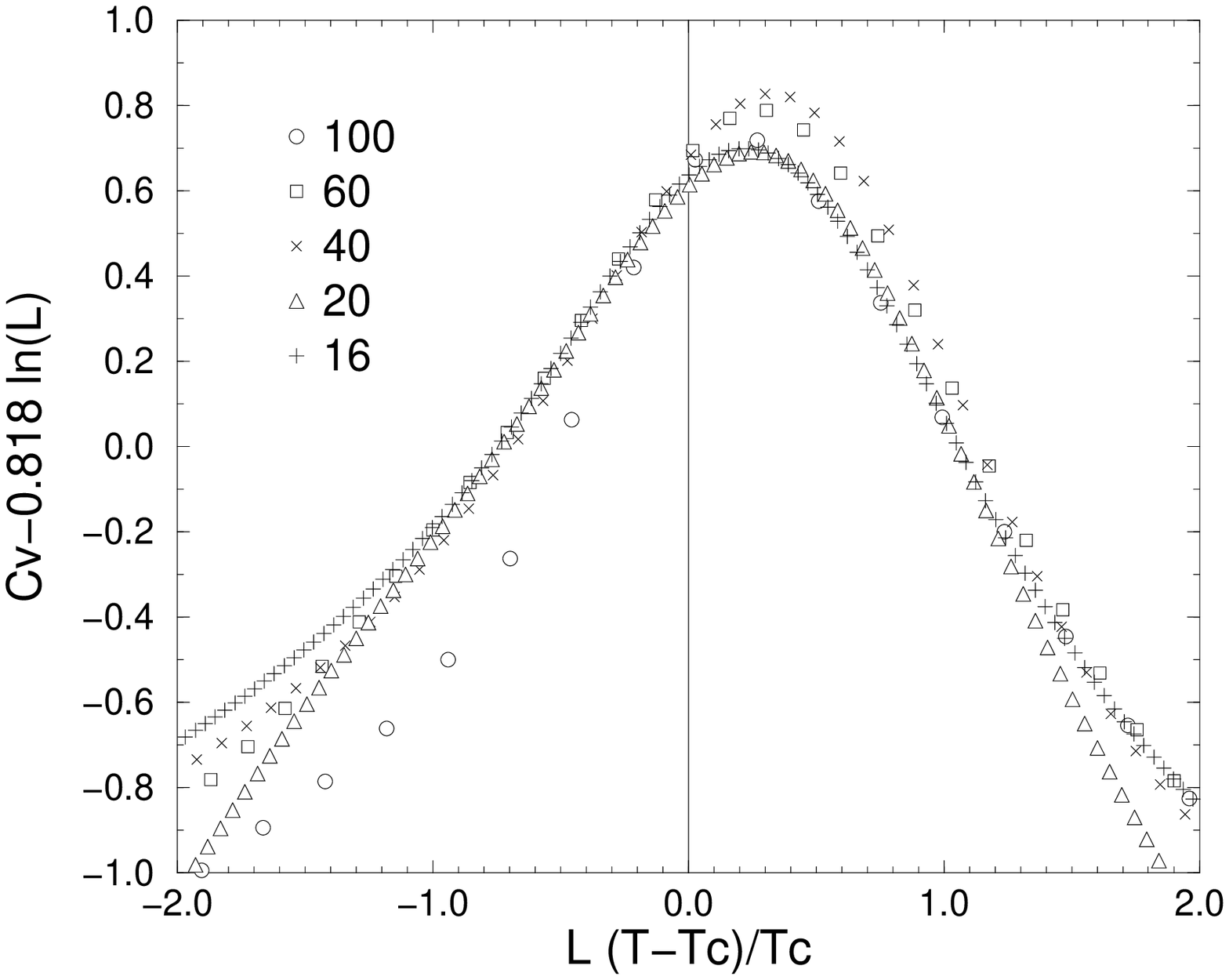,width=19cm,angle=90}}
\end{figure}
\vfill
\centerline{Mu\~noz et. al., Figure 8b}

\newpage
\pagestyle{empty}
\begin{figure}
\centerline{\epsfig{file=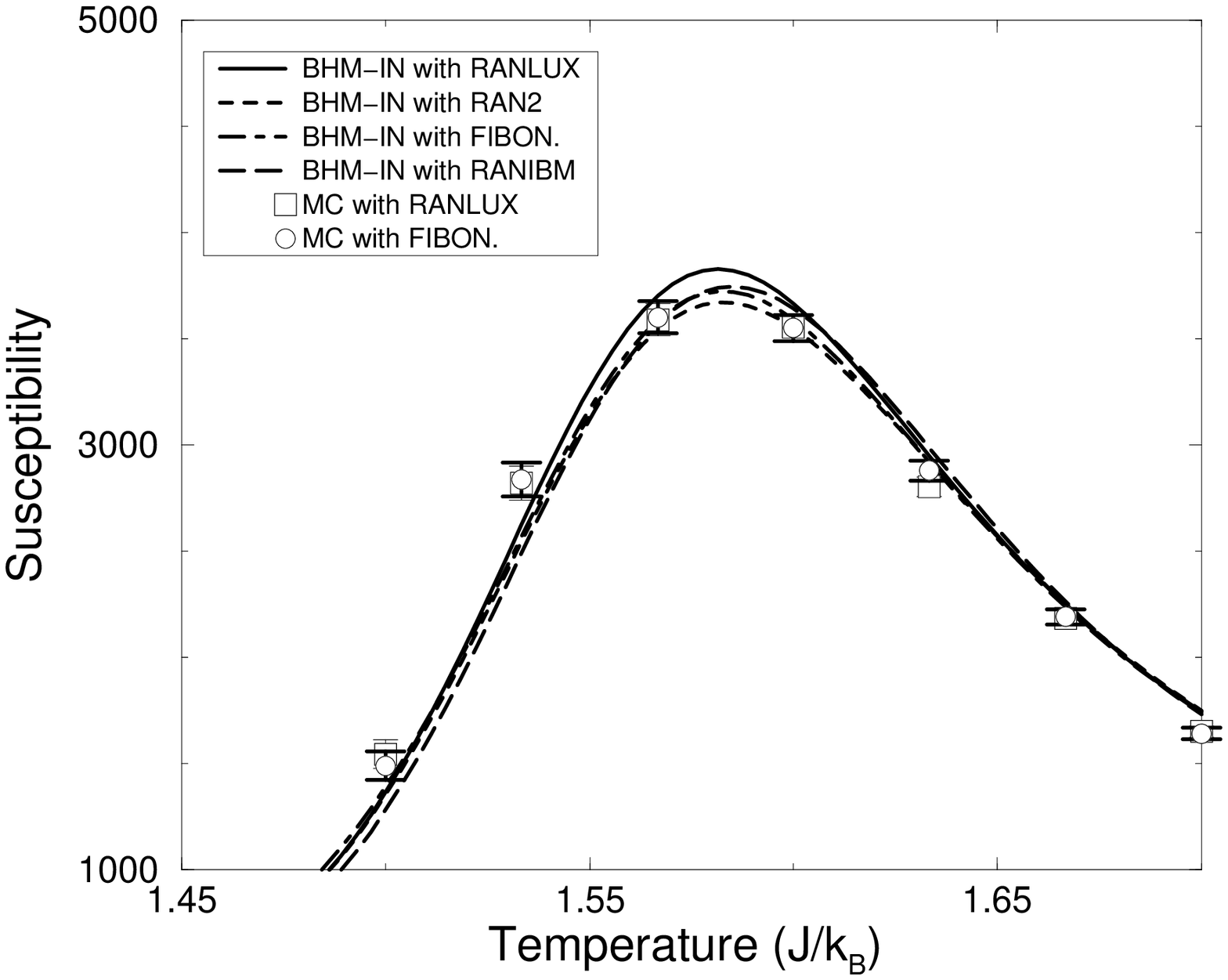,width=19cm,angle=90}}
\end{figure}
\vfill
\centerline{Mu\~noz et. al., Figure 9}

\end{document}